\documentstyle[12pt]{article}

\def\hybrid{\topmargin -20pt    \oddsidemargin 0pt
        \headheight 0pt \headsep 0pt
        \textwidth 6.25in       
        \textheight 9.5in       
        \marginparwidth .875in
        \parskip 5pt plus 1pt   \jot = 1.5ex}
\def\cQ{{\cal Q}}
\def\cG{{\cal G}}
\def\cL{{\cal L}}

\def\cH{{\cal H}}
\def\ket#1{|{#1}\rangle}
\def\noi{\noindent}
\def\half{{1\over2}}
\def\baselinestretch{1.2}

\catcode`\@=11

\def\marginnote#1{}
\def\draftlabel#1{{\@bsphack\if@filesw {\let\thepage\relax
   \xdef\@gtempa{\write\@auxout{\string
      \newlabel{#1}{{\@currentlabel}{\thepage}}}}}\@gtempa
   \if@nobreak \ifvmode\nobreak\fi\fi\fi\@esphack}
        \gdef\@eqnlabel{#1}}
\def\@eqnlabel{}
\def\@vacuum{}
\def\draftmarginnote#1{\marginpar{\raggedright\scriptsize\tt#1}}

\def\draft{\oddsidemargin -.2truein
        \def\@oddfoot{\sl preliminary draft \hfil
        \rm\thepage\hfil\sl\today\quad\militarytime}
        \let\@evenfoot\@oddfoot \overfullrule 3pt
        \let\label=\draftlabel
        \let\marginnote=\draftmarginnote
   \def\@eqnnum{(\theequation)\rlap{\kern\marginparsep\tt\@eqnlabel}%
\global\let\@eqnlabel\@vacuum}  }

\def\preprint{\twocolumn\sloppy\flushbottom\parindent 2em
        \leftmargini 2em\leftmarginv .5em\leftmarginvi .5em
        \oddsidemargin -.5in    \evensidemargin -.5in
        \columnsep .4in \footheight 0pt
        \textwidth 10.in        \topmargin  -.4in
        \headheight 12pt \topskip .4in
        \textheight 6.9in \footskip 0pt
        \def\@oddhead{\thepage\hfil\addtocounter{page}{1}\thepage}
        \let\@evenhead\@oddhead \def\@oddfoot{} \def\@evenfoot{} }

\def\numberbysection{\@addtoreset{equation}{section}
        \def\theequation{\thesection.\arabic{equation}}}

\def\underline#1{\relax\ifmmode\@@underline#1\else
        $\@@underline{\hbox{#1}}$\relax\fi}

\def\titlepage{\@restonecolfalse\if@twocolumn\@restonecoltrue
\onecolumn
     \else \newpage \fi \thispagestyle{empty}\c@page\z@
        \def\thefootnote{\fnsymbol{footnote}} }

\def\endtitlepage{\if@restonecol\twocolumn \else \newpage \fi
        \def\thefootnote{\arabic{footnote}}
        \setcounter{footnote}{0}}  

\catcode`@=12
\relax

\def\figcap{\section*{Figure Captions\markboth
        {FIGURECAPTIONS}{FIGURECAPTIONS}}\list
        {Figure \arabic{enumi}:\hfill}{\settowidth\labelwidth{Figure
999:}
        \leftmargin\labelwidth
        \advance\leftmargin\labelsep\usecounter{enumi}}}
 \relax
\def\tablecap{\section*{Table Captions\markboth
        {TABLECAPTIONS}{TABLECAPTIONS}}\list
        {Table \arabic{enumi}:\hfill}{\settowidth\labelwidth{Table
999:}
        \leftmargin\labelwidth
        \advance\leftmargin\labelsep\usecounter{enumi}}}
 \relax
\def\reflist{\section*{References\markboth
        {REFLIST}{REFLIST}}\list
        {[\arabic{enumi}]\hfill}{\settowidth\labelwidth{[999]}
        \leftmargin\labelwidth
        \advance\leftmargin\labelsep\usecounter{enumi}}}
 \relax
\makeatletter
\newcounter{pubctr}
\def\publist{\@ifnextchar[{\@publist}{\@@publist}}
\def\@publist[#1]{\list
        {[\arabic{pubctr}]\hfill}{\settowidth\labelwidth{[999]}
        \leftmargin\labelwidth
        \advance\leftmargin\labelsep
        \@nmbrlisttrue\def\@listctr{pubctr}
        \setcounter{pubctr}{#1}\addtocounter{pubctr}{-1}}}
\def\@@publist{\list
        {[\arabic{pubctr}]\hfill}{\settowidth\labelwidth{[999]}
        \leftmargin\labelwidth
        \advance\leftmargin\labelsep
        \@nmbrlisttrue\def\@listctr{pubctr}}}
 \relax
\makeatother
\newskip\humongous \humongous=0pt plus 1000pt minus 1000pt

\newif\ifdtup

\font\Scbig=cmss10 scaled\magstep1
\font\Scscr=cmss8 scaled\magstep1
\font\Scscrscr=cmss8
\newfam\Scfam
\textfont\Scfam=\Scbig
\scriptfont\Scfam=\Scscr
\scriptscriptfont\Scfam=\Scscrscr
\def\Sc{\fam\Scfam}

\relax
\hybrid
\def\lvm{\leavevmode\hbox to\parindent{\hfill}}

\def\thefootnote{\fnsymbol{footnote}}
\def\BE{\begin{equation}}
\def\EE{\end{equation}}
\def\BA{\begin{eqnarray}}
\def\EA{\end{eqnarray}}

\def\D{\Delta}

\def\th{\theta}

\def\tt{\bar\tau}

\def\lvm{\leavevmode\hbox to\parindent{\hfill}}
\def\bar{\overline}
\def\req#1{(\ref{#1})}

\def\L{\left}
\def\R{\right}

\def\BE{\begin{equation}}
\def\EE{\end{equation} \vskip 0.30\baselineskip}
\def\BA{\begin{array}}
\def\EA{\end{array}}

\def\noi{\noindent}

\def\frac#1#2{{\textstyle{{#1}\over{#2}}}}
\def\half{{1\over2}}

\def\Kr#1{\delta_{{#1},0}}
\def\ket#1{|{#1}\rangle}

\def\cA{{\cal A}}
\def\cG{{\cal G}}
\def\cH{{\cal H}}

\def\cL{{\cal L}}

\def\cQ{{\cal Q}}

\def\cU{{\cal U}}

\def\open#1{\mbox{{\bf{#1}}}}

\def\oZ{{\open Z}}

\def\ctop{{\Sc c}}
\def\htop{{\Sc h}}

\def\svec{singular vector}

\def\kp{\ket\phi}

\def\ie{{\it i.e.}}

\def\Qz{\cQ_0}
\def\Gz{\cG_0}
\def\Qn{$\Qz$}
\def\Gn{$\Gz$}
\def\kc{{\ket{\chi_T}}}
\def\kct{{\ket{\chi}}}
\def\kcn{{\ket{\chi_{NS}}}}
\def\kcr{{\ket{\chi_{R}}}}
\def\kcc#1#2#3{{\kc_{#1}^{({#2}){#3}}}}

\newif\ifold \oldtrue \def\new{\oldfalse}
\let\ssection=\section
\def\section{\setcounter{equation}{0}\ssection}

\begin{document}
\renewcommand{\theequation}{\thesection.\arabic{equation}}
\newcommand{\beq}{\begin{equation}}
\newcommand{\eeq}[1]{\label{#1}\end{equation}}
\newcommand{\ber}{\begin{eqnarray}}
\newcommand{\eer}[1]{\label{#1}\end{eqnarray}}
\begin{titlepage}
\begin{center}

\hfill IMAFF-FM-01/12,  NIKHEF-01-011\\
\vskip .4in

{\large \bf Construction Formulae for Singular Vectors of the Topological
and of the Ramond N=2 Superconformal Algebras}
\vskip .4in

{\bf Beatriz Gato-Rivera}
\vskip .3in

 {\em Instituto de Matem\'aticas y F\'\i sica Fundamental, CSIC,\\
 Serrano 123, Madrid 28006, Spain} \footnote{e-mail address:
bgato@imaff.cfmac.csic.es}\\
\vskip .2in

{\em NIKHEF-H, Kruislaan 409, NL-1098 SJ Amsterdam, The Netherlands}\\

\vskip 1.0in

\end{center}

\begin{center} {\bf ABSTRACT } \end{center}
\begin{quotation}
We write down one-to-one mappings between the singular vectors of the
Neveu-Schwarz N=2 superconformal algebra and $16 + 16$ types of singular
vectors of the Topological and of the Ramond N=2 superconformal algebras.
As a result one obtains construction formulae for the latter using the
construction formulae for the Neveu-Schwarz singular vectors due to
D\"orrzapf. The indecomposable singular vectors of the Topological and of
the Ramond N=2 algebras (`no-label' and `no-helicity' singular vectors)
cannot be mapped to singular vectors of the Neveu-Schwarz N=2 algebra, but
to {\it subsingular} vectors, for which no construction formulae exist.

\end{quotation}
\vskip 1.5cm

September 2001
\end{titlepage}

\def\baselinestretch{1.2}
\baselineskip 17 pt
\section{Introduction and Notation}\lvm

Singular vectors of infinite dimensional algebras corresponding
to conformal field theories contain an amazing amount
of useful information. They are therefore far from being empty
objects that one simply would like to get rid of.
For example, as a general feature, their decoupling from all other
states in the corresponding Verma module gives rise to differential
equations satisfied by correlators of conformal fields, which can
be solved as a result \cite{BPZ}.
Also, their possible vanishing in the Fock space is
directly connected with the existence of extra states in the Hilbert
space that are not primary and not secondary (not included in any
Verma modules) \cite{MM}. It can even happen that singular vectors
of one theory are directly related to some mathematical structures
of another theory (see for example ref. \cite{BeSe}). For these
reasons it is very convenient to obtain explicit expressions for the
singular vectors of a given algebra. Suitable construction formulae
for these are therefore most helpful.

Regarding the construction of singular vectors,
using either the ``fusion" method or the ``analytic continuation"
method, explicit general expressions have been obtained for the singular
vectors of the Virasoro algebra \cite{VirSV}, the Sl(2)
Kac-Moody algebra \cite{MFF}, the Affine algebra $A_1^{(1)}$
\cite{BaSo}, the N=1 superconformal algebra \cite{N1S}, the
Neveu-Schwarz N=2 superconformal algebra \cite{Doerr1}\cite{Doerr2},
and some W algebras \cite{Walg}. (There is also the method of
construction of singular vertex operators, which produce singular
vectors when acting on the vacuum \cite{KaMa1}).
In some cases it is even possible to transform singular vectors
of an algebra into singular vectors of the same or a different algebra,
simplifying notably the computation of the latter ones. For example,
Kac-Moody singular vectors have been transformed into Virasoro ones,
by using the Knizhnik-Zamolodchikov equation \cite{GP},
and singular vectors of W algebras have been obtained
out of $A_2^{(1)}$ singular vectors via a quantum version of the
highest weight Drinfeld-Sokolov gauge transformations \cite{FGP}.

The N=2 superconformal algebras have appeared in String Theory in
several occasions playing quite different roles. First of all they provide
the symmetries underlying the N=2 strings \cite{Ade}\cite{Marcus}.
These strings seem to be related to
M-theory  since many of the basic objects
of M-theory are realized in the heterotic (2,1) N=2 strings \cite{Marti}.
Second, the N=2 superconformal algebras are encountered also as
the (global) symmetries of the world-sheet of the heterotic string after
compactification from ten to four dimensions preserving N=1 (local)
space-time supersymmetry \cite{HW}. Last but not least, the Topological
N=2 superconformal algebra is realized in the world-sheet of
the bosonic string \cite{BeSe}, as well as in the world-sheet of the
superstrings \cite{BLNW}.

Four years ago the singular vectors of the Topological N=2
algebra were classified \cite{BJI6} taking into account
the relative U(1) charge and the possible annihilation by the
fermionic zero modes of the vector itself and of the primary on
which it is built (BRST-invariance properties). In generic Verma
modules 20 different types of possible \svec s were found whereas
in `no-label' Verma modules, built on indecomposable `no-label'
primaries, 9 types were found (one of them existing only at level
zero). In chiral Verma modules, which are incomplete quotient modules,
the number of different types of \svec s was found to reduce to
just 4. In ref. \cite{BJI6} the whole set of \svec s was
explicitely constructed at level 1, whereas the rigorous
proofs that these types are the only existing ones were
given later in ref. \cite{DB2}. In the latter reference it was
also proved that the Verma modules of the Topological N=2 algebra
are isomorphic to the Verma modules of the Ramond N=2 algebra
and therefore a similar classification of singular vectors hold.
The same is not true for the Verma modules of the Neveu-Schwarz
N=2 algebra, the corresponding singular vectors being in fact less
than half in number than for the other two algebras\footnote{The
Neveu-Schwarz, the Ramond and the Topological N=2 superconformal
algebras are related to each other through the spectral flows
and/or the topological twists. However the Verma modules of the
Neveu-Schwarz N=2 algebra are not isomorphic to the Verma modules of
the other two N=2 algebras and as a consequence its singular vector
structure differs notably from the ones corresponding to the Ramond
and to the Topological N=2 algebras.}.

In this paper we write down one-to-one mappings between
the singular vectors of the Neveu-Schwarz N=2 algebra and
$16 + 16$ types of singular vectors of the Topological and
of the Ramond N=2 algebras. As a result we
obtain construction formulae for the latter (which are absent
in the literature, except for some simple cases) using the
construction formulae for the Neveu-Schwarz singular vectors due to
D\"orrzapf \cite{Doerr1}\cite{Doerr2}. As we will discuss,
the indecomposable singular vectors of the Topological and of the
Ramond N=2 algebras (`no-label' and `no-helicity' singular vectors)
cannot be mapped to singular vectors of the Neveu-Schwarz
N=2 algebra, but only to {\it subsingular} vectors, for which no
construction formulae exist.

The work is organized as follows. In section 2 we review the different
types of \svec s of the N=2 superconformal algebras and we discuss
the basic ingredients to derive the mappings between the \svec s
of the Neveu-Schwarz N=2 algebra and the \svec s of the Topological
and of the Ramond  N=2 algebras. In section 3 we write down these
mappings, which turn into construction formulae for the
\svec s of the Topological and of the Ramond  N=2 algebras
once the \svec s of the Neveu-Schwarz N=2 algebra are expressed
in terms of their construction formulae themselves.
Some final remarks are made in section 4.

\vskip .35in
\noi
{\bf Notation}
\vskip .17in
\noi
{\it Highest weight (h.w.) vectors} denote states annihilated by all
the positive modes of the generators of the algebra:
${\ } \cL_{n > 0} \kct =  \cH_{n > 0} \kct =  {\cG}_{n > 0} \kct
=  {\cQ}_{n > 0} \kct = 0 {\ }$ for the Topological N=2 algebra and
${\ } L_{n > 0} \kct =  H_{n > 0} \kct =  {G^+}_{n > 0} \kct
=  {G^-}_{n > 0} \kct = 0 {\ }$ for the Neveu-Schwarz and for the
Ramond N=2 algebras. These annihilation conditions will be referred
to as the {\it h.w. conditions}.

\noi
{\it Primary states} denote non-singular h.w. vectors.

\noi
{\it Secondary or descendant states} are states obtained by acting on
the h.w. vectors with the negative modes of the generators of the algebra
and with the fermionic zero modes (\Qn\ and \Gn\  for the Topological
N=2 algebra and $G^+_0$ and $G^-_0$ for the Ramond N=2 algebra).
The fermionic zero modes can also interpolate between two h.w.
vectors at the same footing (two primary states or two singular vectors).

\noi
{\it The Verma module} associated to a h.w. vector consists of the
h.w. vector plus the set of secondary states built on it. For most
Verma modules of the Topological and of the Ramond N=2
algebras the h.w. vector is degenerate, the fermionic zero
modes interpolating between the two h.w. vectors.

\noi
{\it Singular vectors} are secondary states that satisfy the h.w.
conditions. As a result they and their descendants decouple from
all other states in the Verma modules and they have zero norm.

\noi
{\it Secondary singular vectors} are \svec s built on singular
vectors. Therefore they cannot `reach back' the singular vectors
on which they are built by acting with the algebra generators.

\noi
{\it Subsingular vectors} are secondary states that satisfy the h.w.
conditions only after setting a \svec\ to zero. Therefore they
become \svec s in quotient modules.

\noi
{\it Topological states} are the states in the Verma modules of the
Topological N=2 algebra.

\noi
{\it R states} are the states in the Verma modules of the
Ramond N=2 algebra.

\noi
{\it NS states} are the states in the Verma modules of the
Neveu-Schwarz N=2 algebra.

\noi
{\it Chiral topological states} $\kc^{G,Q}$ are topological states
annihilated by the two fermionic zero modes $\cG_0$ and $\cQ_0$.

\noi
{\it Chiral R states} $\kcr^{+,-}$ are R states annihilated by
the two fermionic zero modes $G^+_0$ and $G^-_0$.

\noi
{\it Chiral (Antichiral) NS states} $\kcn^{ch}$ ($\kcn^{a}$) are NS
states annihilated by $G^+_{-1/2}$ ($G^-_{-1/2}$).

\noi
{\it $\cG_0$-closed topological states} $\kc^G$ are
non-chiral topological states annihilated by $\cG_0$.

\noi
{\it $\cQ_0$-closed topological states} $\kc^Q$ are
non-chiral topological states annihilated by $\cQ_0$
(they are BRST-invariant since $\cQ_0$ is the BRST charge).

\noi
{\it Helicity $(+)$ R states} $\kcr^+$ are
non-chiral R states annihilated by $G^+_0$.

\noi
{\it Helicity $(-)$ R states} $\kcr^-$ are
non-chiral R states annihilated by $G^-_0$.

\noi
{\it No-label topological states} $\kc$ are
indecomposible topological states that cannot be expressed
as linear combinations of $\cG_0$-closed states,
$\cQ_0$-closed states and chiral states.

\noi
{\it No-helicity R states} $\kcr$ are indecomposible
R states that cannot be expressed as linear combinations of
helicity $(+)$ states, helicity $(-)$ states and chiral states.

\noi
{\it Complete Verma modules} are the Verma modules which
cannot be realized as quotient modules (by setting a singular
vector to zero).

\noi
{\it Generic Verma modules} are complete Verma modules built
on generic h.w. vectors. These are the standard h.w. vectors
annihilated only by the positive modes of the generators of the
algebra and by one (and just one) fermionic zero mode in the case
of the Topological and of the Ramond N=2 algebras.

\noi
{\it No-label (No-helicity) Verma modules} are complete Verma modules
built on no-label topological (no-helicity Ramond) h.w. vectors.

\noi
{\it Chiral Verma modules} are incomplete Verma modules built
on chiral h.w. vectors.



\section{Preliminaries}\lvm

\subsection{N=2 Superconformal algebras}\lvm

The Neveu-Schwarz and the Ramond N=2 superconformal algebras
\cite{Ade}\cite{LVW}\cite{Kir1}\cite{BFK} can be expressed as

\BE\new\BA{lclclcl}
\L[L_m,L_n\R]&=&(m-n)L_{m+n}+{\ctop\over12}(m^3-m)\Kr{m+n}
\,,&\qquad&[H_m,H_n]&=
&{\ctop\over3}m\Kr{m+n}\,,\\
\L[L_m,G_r^\pm
\R]&=&\L({m\over2}-r\R)G_{m+r}^\pm
\,,&\qquad&[H_m,G_r^\pm]&=&\pm G_{m+r}^\pm\,,
\\
\L[L_m,H_n\R]&=&{}-nH_{m+n}\\
\L\{G_r^-,G_s^+\R\}&=&\multicolumn{5}{l}{2L_{r+s}-(r-s)H_{r+s}+
{\ctop\over3}(r^2-\frac{1}{4})
\Kr{r+s}\,,}\EA\label{N2alg}
\EE

\noi
where $L_m$ and $H_m$ are the spin-2 and spin-1 bosonic generators
corresponding to the stress-energy momentum tensor and the U(1) current,
respectively, and
$G_r^+$ and $G_r^-$ are the spin-3/2 fermionic generators. These
are half-integer moded for the case of the Neveu-Schwarz algebra, and
integer moded for the case of the Ramond algebra. The eigenvalues of
the bosonic zero modes $(L_0, \,H_0)$ are the conformal weight and the
U(1) charge of the states. These are split conveniently as
$(\D+l, \,h+q)$ for secondary states, where $l$ and $q$ are the level
and the relative charge of the state and $(\D, \,h)$ are the conformal
weight and U(1) charge of the primary on which the secondary is built.

Observe that we unify the notation for the $U(1)$ charge
of the states of the Neveu-Schwarz algebra and the states of the
Ramond algebra since the $U(1)$ charges of the R states will be
denoted by $h$, instead of $h\pm \half$ which was commonly used in
the past, and their relative charges $q$
are defined to be integer, like for the NS states.

\vskip .17in

The Topological N=2 algebra is the symmetry algebra of  Topological
Conformal Field Theory in two dimensions. It reads \cite{DVV}

\BE\new\BA{lclclcl}
\L[\cL_m,\cL_n\R]&=&(m-n)\cL_{m+n}\,,&\qquad&[\cH_m,\cH_n]&=
&{\ctop\over3}m\Kr{m+n}\,,\\
\L[\cL_m,\cG_n\R]&=&(m-n)\cG_{m+n}\,,&\qquad&[\cH_m,\cG_n]&=&\cG_{m+n}\,,
\\
\L[\cL_m,\cQ_n\R]&=&-n\cQ_{m+n}\,,&\qquad&[\cH_m,\cQ_n]&=&-\cQ_{m+n}\,,\\
\L[\cL_m,\cH_n\R]&=&\multicolumn{5}{l}{-n\cH_{m+n}+{\ctop\over6}(m^2+m)
\Kr{m+n}\,,}\\
\L\{\cG_m,\cQ_n\R\}&=&\multicolumn{5}{l}{2\cL_{m+n}-2n\cH_{m+n}+
{\ctop\over3}(m^2+m)\Kr{m+n}\,,}\EA\qquad m,~n\in\oZ\,.\label{topalgebra}
\EE

\noi
where the fermionic generators $\cQ_m$ and $\cG_m$ correspond
to the spin-1 BRST current and the spin-2 fermionic current, respectively,
\Qn\ being the BRST-charge. The eigenvalues of $(\cL_0, \, \cH_0)$ are
split, as before, as $(\D+l, \, \htop+q)$. This algebra is topological  
becausethe Virasoro generators are BRST-exact: $\cL_m=\half \{\cG_m, \Qz \}$.
This implies, as is well known, that the correlators of the fields do
not depend on the two-dimensional metric.

There is also the twisted N=2 algebra \cite{BFK}\cite{DB4} for which
the generators of the U(1) current are half-integer moded.
As a consequence there is no U(1) charge for this algebra.

\subsection{Relations between the N=2 Superconformal algebras}\lvm

The Neveu-Schwarz, the Ramond and the Topological N=2 superconformal
algebras are connected to each other by the spectral flows and/or
the topological twists. To be precise, the Neveu-Schwarz and the
Topological N=2 algebras are related by the topological twists
whereas the Neveu-Schwarz and the Ramond N=2 algebras are related by
the spectral flows. The twisted N=2 algebra, however, is not
connected to these three N=2 algebras and therefore will not
be considered in what follows.

\subsubsection{Topological twists}\lvm

The Topological N=2 algebra \req{topalgebra} can be obtained
from the Neveu-Schwarz N=2
algebra \req{N2alg} by using one of the two topological twists:

\BE\new\BA{rclcrcl}
\cL^{(\pm)}_m&=&\multicolumn{5}{l}{L_m \pm \half(m+1)H_m\,,}\\
\cH^{(\pm)}_m&=&\pm H_m\,,&{}&{}&{}&{}\\
\cG^{(\pm)}_m&=&G_{m+\half}^{\pm}\,,&\qquad &\cQ_m^{(\pm)}&=
&G^{\mp}_{m-\half} \,,\label{twa}\EA\EE

\noi
These twists, which we denote as $T_W^{\pm }$,
are mirrored under the interchange $H_m \leftrightarrow -H_m$,
${\ } G^{+}_r \leftrightarrow G^{-}_r$. Observe that the h.w.
conditions $G^{\pm}_{1/2}\, \ket{\chi_{NS}} = 0$ of the
Neveu-Schwarz algebra read $\Gz \kc = 0$ under
$T_W^{\pm }$, respectively. Therefore, any NS h.w.
vector results in a \Gn-closed
or chiral topological state, which is also
a h.w. vector as one can easily verify by inspecting the twists
\req{twa}. Conversely, any \Gn-closed or chiral topological
h.w. vector (and only these) transforms under the twists
$T_W^{\pm }$ into a NS h.w. vector.
The zero mode \Qn , in turn, corresponds to the negative modes
$G^{\mp}_{-1/2}$ of the Neveu-Schwarz algebra. Therefore the
topological states
annihilated by \Qn\ become antichiral and chiral NS states
under the twists $T_W^{\pm }$, respectively.

\subsubsection{Spectral flows}\lvm

The even and the odd spectral flows $\,\cU_{\th}$ and $\cA_{\th}$ are
one-parameter families of transformations providing a continuum of
N=2 Superconformal algebras \cite{SS}\cite{LVW}\cite{B1}.
The even spectral flow $\cU_{\th}$ acting on the generators of the
Neveu-Schwarz and of the Ramond N=2 algebras reads

\BE\new\BA{rclcrcl}
\cU_\th \, L_m \, \cU_\th^{-1}&=& L_m
 +\th H_m + {\ctop\over 6} \th^2 \delta_{m,0}\,,\\
\cU_\th H_m \, \cU_\th^{-1}&=&H_m + {\ctop\over3} \th \delta_{m,0}\,,\\
\cU_\th \, G^+_r \, \cU_\th^{-1}&=&G_{r+\th}^+\,,\\
\cU_\th \, G^-_r \, \cU_\th^{-1}&=&G_{r-\th}^-\,,\
\label{spfl} \EA\EE

\noi
satisfying $\, \cU^{-1}_\th = \cU_{(-\th)}$. For $\th=0$ it is just the
identity operator, \ie\ $\, \cU_0={\bf 1}$. It transforms the $(L_0, H_0)$
eigenvalues, \ie\ the conformal weight and the
U(1) charge, $(\Delta, h)$ of a given state as
 $(\Delta-\th h +{\ctop\over6} \th^2, \,h- {\ctop\over3} \th)$. From this
one gets straightforwardly that the level $l$ of any secondary state
changes to $l-\th q$, while the relative charge $q$ remains equal.

The odd spectral flow $\cA_{\th}$ \cite{B1}\cite{BJI4} is given by

\BE\new\BA{rclcrcl}
\cA_\th \, L_m \, \cA_\th^{-1}&=& L_m
 +\th H_m + {\ctop\over 6} \th^2 \delta_{m,0}\,,\\
\cA_\th H_m \, \cA_\th^{-1}&=&- H_m - {\ctop\over3} \th \delta_{m,0}\,,\\
\cA_\th \, G^+_r \, \cA_\th^{-1}&=&G_{r-\th}^-\,,\\
\cA_\th \, G^-_r \, \cA_\th^{-1}&=&G_{r+\th}^+\,,\
\label{ospfl} \EA\EE

\noi
satisfying $\cA_{\th}^{-1} = \cA_{\th}$. It is therefore an involution.
$\cA_{\th}$ is `quasi' mirror symmetric to $\cU_{\th}$,
under the exchange $H_m \to -H_m$, $ G_r^+ \leftrightarrow G_r^-$ and
$\th \to -\th$, and it is in fact the only fundamental spectral flow
since it generates the latter \cite{B1}.
For $\th=0$ it is the mirror map, \ie\ $\cA_0={\cal M}$.
It transforms the $(L_0, H_0)$ eigenvalues of the states as
$(\Delta+\th h +{\ctop\over6} \th^2, - h - {\ctop\over3} \th)$.
The level $l$ of the secondary states changes
to $l + \th q$, while the relative charge $q$ reverses its sign.

For half-integer values of $\th$ the two spectral flows interpolate
between the Neveu-Schwarz algebra and the Ramond algebra.
In particular, for $\th = \pm 1/2$ the NS h.w. vectors
are transformed into R h.w. vectors
algebra with helicities $(\mp)$. As a result the NS
singular vectors are transformed into R singular vectors with
helicities $(\mp)$ built on R primaries with the same helicities.

By performing the topological twists $T_W^{\pm }$ \req{twa}
on the spectral flows \req{spfl} and \req{ospfl} one obtains
the {\it topological} spectral flows \cite{BJI3}\cite{B1}.
Of special importance is the topological odd spectral flow $\cA_1$,
denoted simply as $\cA$, since it transforms all kinds of topological
primary states and \svec s back to topological primary states and
singular vectors. It is given by

\BE\new\BA{rclcrcl}
\cA \, \cL_m \, \cA^{-1}&=& \cL_m - m\cH_m\,,\\
\cA \, \cH_m \, \cA^{-1}&=&-\cH_m - {\ctop\over3} \delta_{m,0}\,,\\
\cA \, \cQ_m \, \cA^{-1}&=&\cG_m\,,\\
\cA {\ } \cG_m \, \cA^{-1}&=&\cQ_m\,.\
\label{autom} \EA\EE

\noi
with  $\cA^{-1} = \cA$. It is therefore an involutive automorphism.
It transforms the $(\cL_0,\cH_0)$ eigenvalues $(\D,\htop)$
of the states as $(\D,-\htop-{\ctop\over3})$, reversing the
relative charge of the secondary states and leaving the level
invariant, as a consequence. In addition, $\cA$ also reverses
the BRST-invariance properties of the states (primary as well as secondary)
mapping \Gn-closed (\Qn-closed) states into \Qn-closed (\Gn-closed)
states, chiral states into chiral states and no-label states into no-label
states. Hence the action of $\cA$ results in the following mappings
between topological singular vectors in different Verma modules:
\begin{eqnarray}
{\cal A}\, \kcc{l,\, \ket{\D,\,\htop}^G}{q}{G} \to \kcc{l,\,
\ket{\D,-\htop-{\ctop\over3}}^Q}{-q}{Q}\ , \ \ \
\ \ {\cal A} \,\kcc{l,\, \ket{\D,\,\htop}^G}{q}{Q} \to \kcc{l,
\, \ket{\D,-\htop-{\ctop\over3}}^Q}{-q}{G} ,\nonumber \\
{\cal A}\, \kcc{l,\, \ket{-l,\,\htop}^G}{q}{G,Q} \to \kcc{l,\,
\ket{-l,-\htop-{\ctop\over3}}^Q}{-q}{G,Q}\ , \ \
\ \ {\cal A} \,\kcc{l,\, \ket{-l,\,\htop}^G}{q}{ } \to \kcc{l,
\, \ket{-l,-\htop-{\ctop\over3}}^Q}{-q}{ },\nonumber\\
{\cal A}\, \kcc{l,\, \ket{0,\,\htop}}{q}{G} \to \kcc{l,\,
\ket{0,-\htop-{\ctop\over3}}}{-q}{Q}\ , \ \ \ \ \ \ \ \
\ \ {\cal A}\, \kcc{l,\, \ket{0,\,\htop}}{q}{Q} \to \kcc{l,\,
\ket{0,-\htop-{\ctop\over3}}}{-q}{G} ,
\label{AADh}
\end{eqnarray}

\noi
and their inverses.

\newpage

\subsection{Singular Vectors of the N=2 Superconformal Algebras}\lvm

In what follows the singular vectors of the Topological, the
Neveu-Schwarz and the Ramond N=2 superconformal algebras
will be denoted as $\kc$, $\kcn$ and $\kcr$, respectively.

\subsubsection{Singular Vectors of the Topological N=2 Algebra}\lvm

{ }From the anticommutator $\{ \cQ_0, \cG_0\} = 2 \cL_0 $ one deduces
\cite{BJI6} that a topological state (primary or secondary)
 with non-zero conformal weight can be either
 \Gn-closed, or \Qn-closed, or a linear combination of both types.
The topological states with zero conformal weight, however, can be
\Qn-closed, or \Gn-closed, or chiral, or no-label (indecomposible).

As a first classification of the topological secondary states one considers
their level $l$, their {\it relative} U(1) charge $q$
and their transformation properties under
\Qn\ and \Gn\ (BRST-invariance properties).
Hence the topological secondary states will be denoted as $\kc_l^{(q)G}$
($\cG_0$-closed), $\kc_l^{(q)Q}$ ($\cQ_0$-closed),
$\kc_l^{(q)G,Q}$ (chiral), and $\kc_l^{(q)}$ (no-label).
The conformal weight and the total U(1) charge of the secondary
states are given by $\D + l$ and $\htop+q$, respectively, where
$\D$ and $\htop$ are the conformal weight and the U(1)
charge of the primary state on which the secondary is built.
Therefore only for $\D + l = 0$ the secondary states can be chiral
or no-label.

There are three different types of topological primaries giving rise
to complete Verma modules: \Gn-closed primaries $\ket{\D,\htop}^G$,
\Qn-closed primaries $\ket{\D,\htop}^Q$, and no-label primaries
$\ket{0,\htop}$. The singular vectors in generic Verma modules, that
is built on \Gn-closed primaries and/or on \Qn-closed primaries
are distributed in the following way \cite{BJI6}\cite{DB2}:

- Ten types built on $\cG_0$-closed primaries
$\ket{\D,\htop}^G$:

\BE
\begin{tabular}{r|l l l l}
{\ }& $q=-2$ & $q=-1$ & $q=0$ & $q=1$\\
\hline\\
\Gn-closed & $-$ & $\kc_l^{(-1)G}$ & $\kc_l^{(0)G}$ & $\kc_l^{(1)G}$\\
\Qn-closed & $\kc_l^{(-2)Q}$ & $\kc_l^{(-1)Q}$ & $\kc_l^{(0)Q}$ & $-$ \\
chiral & $-$ & $\kc_l^{(-1)G,Q}$ &
$\kc_l^{(0)G,Q}$ & $-$ \\
no-label & $-$ & $\kc_l^{(-1)}$ &
$\kc_l^{(0)}$ & $-$\\
\end{tabular}
\label{tabl2}
\EE

- Ten types built on $\cQ_0$-closed primaries
$\ket{\D,\htop}^Q$:

\BE
\begin{tabular}{r|l l l l}
{\ }& $q=-1$ & $q=0$ & $q=1$ & $q=2$\\
\hline\\
$\cG_0$-closed & $-$ & $\kc_l^{(0)G}$ & $\kc_l^{(1)G}$ & $\kc_l^{(2)G}$\\
$\cQ_0$-closed & $\kc_l^{(-1)Q}$ & $\kc_l^{(0)Q}$ & $\kc_l^{(1)Q}$ & $-$ \\
chiral & $-$ & $\kc_l^{(0)G,Q}$ & $\kc_l^{(1)G,Q}$ &
$-$ \\
no-label & $-$ & $\kc_l^{(0)}$ &
$\kc_l^{(1)}$ & $-$\\
\end{tabular}
\label{tabl3}
\EE

\vskip .17in

For $\D \neq 0$ the h.w. vector of the Verma module is degenerate: there
is one \Gn-closed primary state as well as one \Qn-closed primary state,
\Qn\ and \Gn\ interpolating between them.
As a result, for $\D \neq 0$ the singular vectors of table \req{tabl2}
are equivalent to singular vectors of table \req{tabl3} with a shift
on the U(1) charges. That is, the \svec s can be expressed as built
on the \Gn-closed primary or as built on the \Qn-closed primary
(for the details see ref. \cite{BJI6}). In particular, the uncharged
chiral \svec s are equivalent to charged chiral \svec s, provided
the level $l>0$, since for them $\D = - l \neq 0$.
The spectrum of $(\D,\, \htop)$ corresponding to the \Gn-closed primaries
$\ket{\D,\htop}^G$ and to the \Qn-closed primaries $\ket{\D,\htop}^Q$
which contain \svec s in their Verma modules was derived
partially in ref. \cite{BJI6} and completely in ref. \cite{DB3}.

An important observation is that chiral singular vectors
$\kc_{l}^{(q)G,Q}\,$ can be regarded as particular cases of \Gn-closed
singular vectors $\kc_{l}^{(q)G}\,$ and/or as particular cases
of \Qn-closed singular vectors $\kc_{l}^{(q)Q}\,$. That is,
\Gn-closed and \Qn-closed singular vectors may `become' chiral
(although not necessarily) when the conformal weight
of the singular vector turns out to be zero, \ie\ $\D+l=0$. In fact
the singular vectors of types $\kc^{(0)Q}_l$ and $\kc^{(-1)G}_l$
in table \req{tabl2} always become chiral when the conformal
weight is zero and the same is true for
the singular vectors of types $\kc^{(0)G}_l$ and $\kc^{(1)Q}_l$
in table \req{tabl3} \cite{DB2}. In other words, for zero conformal
weight  $\D+l=0$ these types of \svec s are absent from tables
\req{tabl2} and \req{tabl3}.

Inside a given Verma module the topological singular vectors
come in pairs at the same level, one of them \Gn-closed or chiral
(annihilated necessarily by \Gn ) and the other \Qn-closed or chiral
(annihilated necessarily by \Qn ). The only exception to this rule
occurs in the presence of no-label \svec s. In this `degenerate' case
there are four \svec s at the same level: one \Gn-closed, one \Qn-closed,
one chiral and one no-label. In the general case there are only two
\svec s at the same level, the fermionic zero modes \Gn\ and \Qn\
interpolating  between them, although it can also happen that both are
chiral and then disconnected from each other \cite{DB3}. To be precise,
\Gn\ and \Qn\ interpolate between two singular vectors
with non-zero conformal weight in both directions, whereas they
produce secondary singular vectors when acting on singular vectors
with zero conformal weight. That is,
inside a given Verma module $V(\Delta,\htop)$ and for a given
level $l$ the topological singular vectors with non-zero conformal
weight are connected by the action of \Qn\ and \Gn\ as:
\begin{eqnarray}  \Qz \, \kcc{l}{q}{G} \to \kcc{l}{q-1}{Q} ,
 \qquad\Gz \,\kcc{l}{q}{Q} \to \kcc{l}{q+1}{G}\, , \label{GQh}
\end{eqnarray}
\noi
where the arrows can be reversed (up to constants), using \Gn\ and
\Qn\ respectively, since the conformal weight of the singular vectors
is different from zero, \ie\ $\D+l\neq0$. Otherwise, on the right-hand side
of the arrows one obtains {\it chiral secondary} singular vectors
which cannot `come back' to the \svec s on the left-hand side:
\begin{eqnarray}  \Qz \,\kcc{l=-\D}{q}{G} \to \kcc{l=-\D}{q-1}{G,Q} ,
 \qquad\Gz \,\kcc{l=-\D}{q}{Q} \to \kcc{l=-\D}{q+1}{G,Q}\, . \label{GQch}
\end{eqnarray}

Regarding no-label singular vectors
$\kc_l^{(q)}$, they always satisfy $\D+l=0$. The action of \Gn\ and \Qn\
on a no-label singular vector produce three secondary singular vectors
which cannot come back to the no-label
\svec $\kcc{l=-\D}{q}{ }$ by acting with \Gn\ and \Qn\ :
 \begin{eqnarray} \Qz \,\kcc{l=-\D}{q}{ } \to \kcc{l=-\D}{q-1}{Q},
 \ \Gz \,\kcc{l=-\D}{q}{ } \to \kcc{l=-\D}{q+1}{G}, \
 \Gz \, \Qz \,\kcc{l=-\D}{q}{ } \to \kcc{l=-\D}{q}{G,Q}. \label {QGnh}
 \end{eqnarray}

The \Qn-closed and no-label h.w. vectors (primaries or singular vectors)
are transformed, under the twists $T_W^{\pm }$ \req{twa}, into states
of the Neveu-Schwarz algebra which are not h.w. vectors since they are not
annihilated by one of the modes $G^{\pm}_{1/2}$ (because the topological
state is not annihilated by \Gn ). The \Gn-closed and chiral h.w. vectors,
however, are transformed under $T_W^{\pm }$ into NS h.w. vectors.
In particular the \Gn-closed primaries $\ket{\D, \,h}^G$
generate the topological Verma modules which are mapped to
the complete NS Verma modules, whereas the chiral primaries
$\ket{0, \, h}^{G,Q}$ generate topological chiral (incomplete) Verma
modules which are mapped to antichiral and chiral (incomplete)
NS Verma modules under $T_W^{\pm }$, respectively.

We will see in next section that, with the exception of the
indecomposable no-label \svec s,
all other 16 types of topological \svec s in tables \req{tabl2} and
\req{tabl3} can be mapped to NS \svec s using the twists $T_W^{\pm }$
and the involutive automorphism $\cA$ \req{autom}.  The no-label \svec s,
however, can be mapped only to NS {\it subsingular} vectors \cite{BJI5},
as was shown in ref. \cite{DB1}, for which there are no construction
formulae.

Regarding no-label Verma modules, the spectrum of $\htop$ corresponding
to the no-label primaries $\ket{0,\htop}$ containing \svec s in
their Verma modules was derived in ref. \cite{DB3}.
The possible existing types of singular vectors in no-label
Verma modules are the following nine types
\cite{BJI6}\cite{DB2}\cite{DB3} built on no-label primaries
$\ket{0,\htop}$:

\BE
\begin{tabular}{r|l l l l l}
{\ }& $q=-2$ & $q=-1$ & $q=0$ & $q=1$ & $q=2$\\
\hline\\
$\cG_0$-closed & $-$ & $\kc_l^{(-1)G}$ & $\kc_l^{(0)G}$ & $\kc_l^{(1)G}$ &
$\kc_l^{(2)G}$\\
$\cQ_0$-closed & $\kc_l^{(-2)Q}$ &$\kc_l^{(-1)Q}$ &
$\kc_l^{(0)Q}$ & $\kc_l^{(1)Q}$ & $-$ \\
chiral & $-$ & $-$ & $\kc_0^{(0)G,Q}$ & $-$ & $-$ \\
\end{tabular}
\label{tabl4}
\EE
\vskip .17in

Since the no-label primaries do not have a counterpart
in the Neveu-Schwarz algebra, the \svec s built on them
cannot be mapped to NS \svec s unless they are located
in the submodules generated by the level-zero \svec s
$\cG_0\ket{0,\htop} = \ket{0,\htop+1}^G$ or
$\cQ_0\ket{0,\htop} = \ket{0,\htop-1}^Q$. In these cases the \svec s
can be regarded as built on the \Gn-closed h.w. vector
$\ket{0,\htop+1}^G$ or on the \Qn-closed h.w. vector $\ket{0,\htop-1}^Q$
and they fit into the tables \req{tabl2} and \req{tabl3}, respectively.
Observe that all the \svec s in no-label Verma modules are either
\Gn-closed or \Qn-closed, except a unique chiral \svec\
which is at level zero: $\cG_0\cQ_0\ket{0,\htop}$.
There are no no-label \svec s built on no-label primaries because
their level, that is their conformal weight, must be zero and it is not  
possible to construct a no-label \svec\ at level zero just by the
action of the fermionic zero modes \Gn\ and \Qn .

\subsubsection{Singular Vectors of the Ramond N=2 Algebra}\lvm

The fermionic zero modes characterize the R states as being annihilated
by $G_0^+$ or by $G_0^-$ (or as a linear combination of these), as the
anticommutator $\{G_0^+,G_0^-\} = 2L_0 - {\ctop\over12}{\ }$ shows.
We call these states `helicity' $(+)$ and `helicity' $(-)$ states.
However, for conformal weight $\Delta = {\ctop\over24}$ the R states
can be annihilated by both $G_0^+$ and $G_0^-$ (we call them chiral),
and there also exist indecomposable `no-helicity' states which cannot
be expressed as linear combinations of  helicity $(+)$, helicity $(-)$
and chiral states. These no-helicity states, \svec s in particular,
have been reported for the first time in ref. \cite{DB1},
and they were completely overlooked in the early literature.
The R singular vectors will be denoted therefore as $\ket{\chi_R}_l^{(q)+}$,
$\ket{\chi_R}_l^{(q)-}$, $\ket{\chi_R}_l^{(q)+-}$ or $\ket{\chi_R}_l^{(q)}$,
where, in addition to the level $l$ and the relative charge $q$,
the helicities indicate that the vector is annihilated by
$G_0^+$ or $G_0^-$, or both, or none, respectively.

We already see that the case $\D = {c\over 24}$ for the Ramond N=2 algebra
is very similar to the case $\D = 0$ for the Topological N=2 algebra,
the chiral $(+-)$ and no-helicity R states being analogous to the
chiral $(G,Q)$ and no-label topological states, respectively. This is
not just a similarity but a consequence of the fact that
the Verma modules of the Ramond N=2 algebra are isomorphic to the
Verma modules of the Topological N=2 algebra \cite{DB3}.
In particular one can construct a one-to-one map between each
R \svec\ and a topological \svec\ at the same level and with the same
relative charge. As a consequence, most results concerning the
topological \svec s, in particular the classification summarized in
tables \req{tabl2}, \req{tabl3} and \req{tabl4}, can be transferred
straightforwardly to the R \svec s. The argument goes as follows \cite{DB3}.

Let us compose the topological twists
\req{twa}, which transform the Topological N=2 algebra into the
Neveu-Schwarz N=2 algebra, with the spectral flows \req{spfl} and
\req{ospfl}, which transform the Neveu-Schwarz N=2 algebra
into the Ramond N=2 algebra.
By analysing all possible combinations one obtains
that the only map that transforms the topological states into R
states, preserving the level $l$ and the relative charge $q$, is given
by the exactly equivalent compositions $\cA_{-1/2}\, (T_W^-)^{-1}$ and
$\cU_{-1/2}\,(T_W^+)^{-1}$:
\BE  \ket{\chi_R}^{(q)}_l = \cA_{-1/2}\, (T_W^-)^{-1} \kc^{(q)}_l =
 \cU_{-1/2}\,(T_W^+)^{-1}  \kc^{(q)}_l  \,.
\label{compo} \EE
\noi
This map is one-to-one because it transforms
every topological state into a R state, and the other way around:
$\kc^{(q)}_l =  (T_W^-) \, \cA_{-1/2}\, \ket{\chi_R}^{(q)}_l =
(T_W^+) \, \cU_{1/2}\, \ket{\chi_R}^{(q)}_l \,$ (from now on we
will keep only the composition $\cA_{-1/2}\, (T_W^-)^{-1}$).
Furthermore if $\kc^{(q)}_l$ is singular, i.e. satisfies the
topological h.w. conditions
${\ } \cL_{n \geq 1} \kc^{(q)}_l =  \cH_{n \geq 1} \kc^{(q)}_l =
{\cG}_{n \geq 1} \kc^{(q)}_l =  {\cQ}_{n \geq 1} \kc^{(q)}_l = 0 {\ }$,
then $\ket{\chi_R}^{(q)}_l$ is also singular, satisfying in turn
the R h.w. conditions
${\ } L_{n \geq 1} \ket{\chi_R}^{(q)}_l =
H_{n \geq 1} \ket{\chi_R}^{(q)}_l =
{G^+}_{n \geq 1} \ket{\chi_R}^{(q)}_l =
{G^-}_{n \geq 1} \ket{\chi_R}^{(q)}_l = 0 {\ }$. To see this one has to
study first the transformation, under $\cA_{-1/2}\, (T_W^-)^{-1}$, of the
h.w. vectors of the topological Verma modules. There are four cases
to analyse, corresponding to the topological h.w. vectors being \Gn-closed
$\ket{\D, \htop}^G$, \Qn-closed $\ket{\D, \htop}^Q$, chiral
$\ket{0, \htop}^{G,Q}$, and no-label $\ket{0, \htop}$. By carefully keeping
track of the transformation of the positive and zero modes of the
topological generators one obtains that under $\cA_{-1/2}\, (T_W^-)^{-1}$:

\vskip .2in

i) The \Gn-closed topological h.w. vectors $\ket{\D, \htop}^G$ are
mapped to helicity-$(+)$ R h.w. vectors
$\ket{\D_R, \htop_R}^+_R=\ket{\D+{\ctop\over24},\htop+{\ctop\over6}}^+_R$.

ii) The \Qn-closed topological h.w. vectors $\ket{\D, \htop}^Q$ are
mapped to helicity-$(-)$ R h.w. vectors
$\ket{\D_R, \htop_R}^-_R=\ket{\D+{\ctop\over24},\htop+{\ctop\over6}}^-_R$.

iii) The chiral topological h.w. vectors $\ket{0, \htop}^{G,Q}$ are
mapped to chiral R h.w. vectors $\ket{\D_R, \htop_R}^{+-}_R=
\ket{{\ctop\over24},\htop+{\ctop\over6}}^{+-}_R$.

iv) The no-label topological h.w. vectors $\ket{0, \htop}$ are
mapped to no-helicity R h.w. vectors
$\ket{\D_R, \htop_R}_R=\ket{{\ctop\over24},\htop+{\ctop\over6}}_R$.

\vskip .2in
Now by taking into account that \svec s are just particular cases of
h.w. vectors one deduces that the topological \svec s are
transformed under $\cA_{-1/2}\, (T_W^-)^{-1}$ into R \svec s at the
same level $l$, with the same relative charge $q$, and
with the helicities determined by the exchange
$\, G \to + \,, Q \to - \,$. That is,
$\kc^{(q)G}_l \to \ket{\chi_R}_l^{(q)+}$,
$\kc^{(q)Q}_l \to \ket{\chi_R}_l^{(q)-}$,
$\kc^{(q)G,Q}_l \to \ket{\chi_R}_l^{(q)+-}$ and
$\kc^{(q)}_l \to \ket{\chi_R}_l^{(q)}$.
Therefore the classification of the
topological singular vectors in tables
\req{tabl2}, \req{tabl3} and \req{tabl4}, is also valid for the R
\svec s, simply by taking into account
that $\D_R=\D+{\ctop\over24}$, $\htop_R=\htop+{\ctop\over6}$, and
exchanging the labels $G \to +\,$ and $\,Q \to -\,$. As a consequence
the R \svec s in complete Verma modules are distributed as follows:

- Ten types built on helicity $(+)$ primaries
$\ket{\D,\htop}^+$:

\BE
\begin{tabular}{r|l l l l}
{\ }& $q=-2$ & $q=-1$ & $q=0$ & $q=1$\\
\hline\\
helicity $(+)$ & $-$ & $\kcr_l^{(-1)+}$ & $\kcr_l^{(0)+}$ & $\kcr_l^{(1)+}$\\
helicity $(-)$ & $\kcr_l^{(-2)-}$ & $\kcr_l^{(-1)-}$ & $\kcr_l^{(0)-}$ & $-$ \\
chiral & $-$ & $\kcr_l^{(-1)+,-}$ &
$\kcr_l^{(0)+,-}$ & $-$ \\
no-helicity & $-$ & $\kcr_l^{(-1)}$ &
$\kcr_l^{(0)}$ & $-$\\
\end{tabular}
\label{tabl5}
\EE

- Ten types built on helicity $(-)$ primaries
$\ket{\D,\htop}^-$:

\BE
\begin{tabular}{r|l l l l}
{\ }& $q=-1$ & $q=0$ & $q=1$ & $q=2$\\
\hline\\
helicity $(+)$ & $-$ & $\kcr_l^{(0)+}$ & $\kcr_l^{(1)+}$ & $\kcr_l^{(2)+}$\\
helicity $(-)$ & $\kcr_l^{(-1)-}$ & $\kcr_l^{(0)-}$ & $\kcr_l^{(1)-}$ & $-$ \\
chiral & $-$ & $\kcr_l^{(0)+,-}$ & $\kcr_l^{(1)+,-}$ &
$-$ \\
no-helicity & $-$ & $\kcr_l^{(0)}$ &
$\kcr_l^{(1)}$ & $-$\\
\end{tabular}
\label{tabl6}
\EE

- Nine types built on no-helicity primaries $\ket{{\ctop\over24},\htop}$:

\BE
\begin{tabular}{r|l l l l l}
{\ }& $q=-2$ & $q=-1$ & $q=0$ & $q=1$ & $q=2$\\
\hline\\
helicity $(+)$ & $-$ & $\kcr_l^{(-1)+}$ & $\kcr_l^{(0)+}$ & $\kcr_l^{(1)+}$ &
$\kcr_l^{(2)+}$\\
helicity $(-)$ & $\kcr_l^{(-2)-}$ &$\kcr_l^{(-1)-}$ &
$\kcr_l^{(0)-}$ & $\kcr_l^{(1)-}$ & $-$ \\
chiral & $-$ & $-$ & $\kcr_0^{(0)+,-}$ & $-$ & $-$ \\
\end{tabular}
\label{tabl7}
\EE
\vskip .17in

Analogously as for the topological case, for $\D \neq {\ctop\over24}$
the h.w. vector of the generic Verma modules
is degenerate: there is one helicity $(+)$ primary state
as well as one helicity $(-)$ primary state, $G_0^+$ and $G_0^-$
interpolating between them. In addition, for $\D+l={\ctop\over24}$ some
types of \svec s are absent from tables \req{tabl5} and \req{tabl6}:
the singular vectors of types $\kcr^{(0)-}_l$ and $\kcr^{(-1)+}_l$
in table \req{tabl5} and the singular vectors of types
$\kcr^{(0)+}_l$ and $\kcr^{(1)-}_l$ in table \req{tabl6} `become' chiral.
Inside a Verma module $G_0^+$ and $G_0^-$ interpolate between two
singular vectors, in both directions if $\D+l \neq {\ctop\over24}$,
or in only one direction producing secondary \svec s if
$\D+l = {\ctop\over24}$. Therefore the R \svec s come in pairs at the
same level, like the topological ones. Generically one of the \svec s
belong to the `$(+)$ sector' and the other to the `$(-)$ sector', as was
stated in the early literature, although there are many exceptions
(see the discussion in ref. \cite{DB3}). For example, the no-helicity
\svec s do not belong to any of these sectors.

The spectrum of $(\D,\, \htop)$ corresponding to the helicity $(+)$
primaries $\ket{\D,\htop}^+$ and to the helicity $(-)$
primaries $\ket{\D,\htop}^-$ which contain \svec s
in their Verma modules was written down for the first time in ref.
\cite{BFK}. The existence of no-helicity \svec s was overlooked,
however. The spectrum of $\htop$
corresponding to the no-helicity primaries $\ket{{\ctop\over24},\htop}$
containing \svec s in their Verma modules was written in ref. \cite{DB3}.

In the case of the Ramond N=2 algebra both the
helicity $(+)$ and the helicity $(-)$ h.w. vectors
are transformed into NS h.w. vectors by the spectral flows
$\cU_{\th}$ and $\cA_{\th}$  (for suitable values of $\th$). This
contrasts the situation for the case of the Topological N=2 algebra
where the \Qn-closed h.w. vectors are not mapped to NS h.w. vectors
by the topological twists. The chiral R h.w. vectors in turn are
transformed into chiral and antichiral NS h.w. vectors too.
The no-helicity h.w. vectors (primaries or singular vectors) are
not transformed into NS h.w. vectors under the spectral flows
$\cU_{\th}$ and $\cA_{\th}$, for any value of $\th$.
The no-helicity \svec s can be mapped only to NS {\it subsingular}
vectors \cite{DB1}, as was the case for the no-label topological
\svec s. All other 16 types of R \svec s in tables \req{tabl5} and
\req{tabl6} can be mapped to NS \svec s using the spectral flows.
The \svec s built on no-helicity h.w. vectors can be mapped to NS
\svec s only if they are descendants of the level-zero \svec s
$\ G_0^+\,\ket{{\ctop\over24},\htop} = \ket{{\ctop\over24},\htop+1}^+$
or $\ G_0^-\,\ket{{\ctop\over24},\htop} = \ket{{\ctop\over24},\htop-1}^-$
since in these cases they fit in tables \req{tabl5} and \req{tabl6},
respectively.

\subsubsection{Singular Vectors of the Neveu-Schwarz N=2 Algebra}\lvm

The Neveu-Schwarz N=2 algebra contains no fermionic zero modes.
As a consequence the classification of h.w. vectors, \svec s and
Verma modules is much simpler for this algebra than for the
Topological and for the Ramond N=2 algebras. Basically one
considers only generic and chiral (and antichiral) h.w. vectors
and \svec s. The chiral and antichiral h.w. vectors and \svec s are
annihilated by the negative modes $G^+_{-1/2}$ and $G^-_{-1/2}$,
respectively. The generic h.w. vectors give rise to complete
Verma modules while the chiral and antichiral h.w. vectors
produce incomplete chiral Verma modules. In the generic Verma
modules the possible \svec s belong to the following types
\cite{Doerr2}\cite{DB2}:

\BE
\begin{tabular}{r| l l l}
{\ }&  $q=-1$ & $q=0$ & $q=1$\\
\hline\\
generic & $\kcn_l^{(-1)}$ & $\kcn_l^{(0)}$ & $\kcn_l^{(1)}$\\
chiral & $-$ &
$\kcn_l^{(0)ch}$ & $\kcn_l^{(1)ch}$ \\
antichiral & $\kcn_l^{(-1)a}$ &
$\kcn_l^{(0)a}$ & $-$ \\
\end{tabular}
\label{tabl8}
\EE

The spectrum of $(\D,\, \htop)$ corresponding to the generic
NS primaries $\ket{\D,\, \htop}$
which contain \svec s in their Verma modules was obtained
for the first time in refs. \cite{BFK} and \cite{Nam}. The fact
that $|q| \leq 1$ for all NS \svec s, not only for the primitive
\svec s but also for the secondary ones, was only realized and proved
in ref. \cite{Doerr2}, however.

All the h.w. vectors
and \svec s of the Neveu-Schwarz N=2 Algebra can be mapped to h.w.
vectors and \svec s of the Topological and of the Ramond N=2 algebras.
The converse is not true, however, as we have already pointed out.

\section{Construction Formulae}\lvm

About six years ago construction formulae for the singular vectors of
the Neveu-Schwarz N=2 algebra were computed by D\"orrzapf.
Using the fusion method explicit formulae were obtained
\cite{Doerr1} for all the charged singular vectors, and for
a class of uncharged singular vectors. Later using the analytic
continuation method explicit formulae were obtained
for all the uncharged singular vectors \cite{Doerr2}.
In what follows we will show that these construction formulae for the
NS singular vectors also provide construction formulae for 16 types
of topological singular vectors and for 16 types of R \svec s, as
one can write maps from the NS \svec s to these $16 + 16$ types of
topological and R \svec s.

\subsection{Maps from NS to topological \svec s}\lvm

To derive the maps from the NS \svec s to the topological ones
we will proceed in the following way. First we will construct
the `box' diagrams obtained by the actions of the fermionic
zero modes \Gn, \Qn , eqns. \req{GQh}-\req{GQch}, and the
action of the spectral flow automorphism $\cA$, eqns. \req{AADh}.
We will be interested only in the box diagrams which contain \Gn-closed
topological singular vectors built on \Gn-closed primaries; \ie\ the
\svec s of type $\kc^{(q)G}_{l,{\kp^G}}{\ }$ in table \req{tabl2},
since only these types have a direct relation with the
generic NS singular vectors via the topological twists. From
the box diagrams one can deduce straightforwardly two different
maps from the NS \svec s to each topological singular vector,
taking into account that in every box diagram the topological \svec\ of
type $\kc^{(q)G}_{l,{\kp^G}}{\ }$ can be transformed into two NS \svec s
using the two topological twists $T_W^{\pm }$ \req{twa}.
These two NS \svec s are mirror-symmetric under
the exchange $H_m \to -H_m$ and $G^+_r \leftrightarrow G^-_r$; therefore
they have opposite U(1) charges and are located in mirror-symmetric
Verma modules:
\BE \kc^{(q)G}_{l,\,{\ket{\D,\,\htop}^G}} =
 T_W^+ {\ } \ket{\chi_{NS}}^{(q)}_{l-q/2,\,{\ket{\D-\htop/2,\,\htop}}}
 = T_W^- {\ }
 \ket{\chi_{NS}}^{(-q)}_{l-q/2,\,{\ket{\D-\htop/2,-\htop}}}\,.
 \label{mtq} \EE

Let us start with the box diagrams which contain an uncharged singular
vector of type $\kc^{(0)G}_{l,{\kp^G}}{\ }$ in the Verma module
$V(\ket{\D,\htop}^G)$. For non-zero conformal weight, $\D+l \neq 0$, the
box diagram, shown in \req{diab0}, consists of
singular vectors of types $\kc^{(0)G}_{l,{\kp^G}}{\ }$ and
$\kc^{(-1)Q}_{l,{\kp^G}}{\ }$, at level $l$ in the Verma module
$V(\ket{\D,\htop}^G)$, and singular
vectors of types $\kc^{(0)Q}_{l,{\kp^Q}}{\ }$
and $\kc^{(1)G}_{l,{\kp^Q}}{\ }$, also at level $l$ in the Verma module
 $V(\ket{\D,-\htop-\ctop/3}^Q)$.

\vskip .15in

\def\btggo  {\mbox{$\kc_{l,\, \ket{\D,\,\htop}^G}^{(0)G} $}}
\def\btqqo  {\mbox{$\kc_{l,\, \ket{\D,-\htop-{\ctop\over3}}^Q}^{(0)Q} $}}
\def\btqqm  {\mbox{$\kc_{l,\, \ket{\D,\,\htop}^G}^{(-1)Q} $}}
\def\btggp  {\mbox{$\kc_{l,\, \ket{\D,-\htop-{\ctop\over3}}^Q}^{(1)G} $}}

  \begin{equation}
  \begin{array}{rcl}
   \btggo &
  \stackrel{\Qz}{\mbox{------}\!\!\!\longrightarrow}
  & \btqqm \\[3 mm]
   \cA\,\updownarrow\ && \ \updownarrow\, \cA
  \\[3 mm]   \btqqo \! & \stackrel{\Gz}
  {\mbox{------}\!\!\!\longrightarrow} & \! \btggp
  \end{array} \label{diab0} \end{equation}

\vskip .15in

\noi
The arrows \Gn\ and \Qn\ can be reversed (up to constants) using \Qn\
and \Gn\ respectively; that is, the fermionic zero modes interpolate
between two singular vectors, one charged and one uncharged,
at the same level in the same Verma module.

For $\D+l = 0$, the conformal weight of the singular
vectors is zero, so that the corresponding
arrows $\Qz$, $\Gz$ cannot be reversed, producing
{\it secondary} chiral singular vectors
$\kc_{l,\,\ket{-l,\,\htop}^G}^{(-1)G,Q}$ and
$\kc_{l,\,\ket{-l,-\htop-\ctop/3}^Q}^{(1)G,Q}$ on the right-hand side,
at level zero with respect to the singular vectors on the left-hand side.
The corresponding box diagram is therefore:

\vskip .15in

\def\btggoc  {\mbox{$\kc_{l,\, \ket{-l,\,\htop}^G}^{(0)G} $}}
\def\btqqoc  {\mbox{$\kc_{l,\, \ket{-l,-\htop-{\ctop\over3}}^Q}^{(0)Q} $}}
\def\btqqmc  {\mbox{$\kc_{l,\, \ket{-l,\,\htop}^G}^{(-1)G,Q} $}}
\def\btggpc  {\mbox{$\kc_{l,\, \ket{-l,-\htop-{\ctop\over3}}^Q}^{(1)G,Q} $}}

  \begin{equation}
  \begin{array}{rcl}
   \btggoc &
  \stackrel{\Qz}{\mbox{------}\!\!\!\longrightarrow}
  & \btqqmc \\[3 mm]
   \cA\,\updownarrow\ && \ \updownarrow\, \cA
  \\[3 mm]   \btqqoc \! & \stackrel{\Gz}
  {\mbox{------}\!\!\!\longrightarrow} & \! \btggpc
  \end{array} \label{diab0ch} \end{equation}

\vskip .15in

The untwisting of the uncharged \svec\
$\kc^{(0)G}_{l,\,{\ket{\D,\,\htop}^G}}{\ }$, using $T_W^{\pm }$ \req{twa},
produces two uncharged mirror-symmetric NS \svec s
located in mirror-symmetric Verma modules. Therefore,
as shown in \req{mtq}, the twisting
of two uncharged mirror-symmetric NS \svec s, using $T_W^+$ and
$T_W^-$ respectively, produces the same uncharged topological \svec\ of
type $\kc^{(0)G}_{l,{\kp^G}}{\ }$. That is, one has the maps:
\BE \kc^{(0)G}_{l,\,{\ket{\D+\htop/2,\,\htop}^G}} =
 T_W^+ {\ } \ket{\chi_{NS}}^{(0)}_{l,\,{\ket{\D,\,\htop}}}
 = T_W^- {\ }
 \ket{\chi_{NS}}^{(0)}_{l,\,{\ket{\D,-\htop}}}\,,  \label{mtu} \EE
\noi
 where we have redefined $\D$ as the conformal weight of the NS primaries.
 The maps from the NS \svec s to the remaining topological \svec s
 in diagrams \req{diab0} and \req{diab0ch} can be derived now
 resulting in the following expressions:

\BE \begin{array}{lcl}

  {\ }\kc^{(0)Q}_{l,\,{\ket{\D+\htop/2,-\htop-{\ctop\over3}}^Q}} &=&
\cA {\ } T_W^+ {\ } \ket{\chi_{NS}}^{(0)}_{l,\,{\ket{\D,\,\htop}}} \\

  {\ }\kc^{(-1)Q}_{l,\,{\ket{\D+\htop/2,\,\htop}^G}} &=&
\Qz {\ } T_W^+ {\ } \ket{\chi_{NS}}^{(0)}_{l,\,{\ket{\D,\,\htop}}} \\

  {\ }\kc^{(1)G}_{l,\,{\ket{\D+\htop/2,-\htop-{\ctop\over3}}^Q}} &=& \cA
 {\ }\Qz {\ } T_W^+ {\ } \ket{\chi_{NS}}^{(0)}_{l,\,{\ket{\D,\,\htop}}} \\

  {\ }\kc^{(-1)G,Q}_{l,\,{\ket{-l,\,\htop}^G}} &=& \Qz {\ }
T_W^+ {\ } \ket{\chi_{NS}}^{(0)}_{l,\,{\ket{-l-\htop/2,\,\htop}}} \\

  {\ }\kc^{(1)G,Q}_{l,\,{\ket{-l,-\htop-{\ctop\over3}}^Q}} &=& \cA {\ }\Qz
  {\ } T_W^+ {\ } \ket{\chi_{NS}}^{(0)}_{l,\,{\ket{-l-\htop/2,\,\htop}}}
 \label{cfu} \end{array} \EE
\noi
and mirror-symmetric expressions using
$T_W^- \,\ket{\chi_{NS}}^{(0)}_{l,\,{\ket{\D,-\htop}}} $. Observe that
the last two maps, to chiral \svec s, are not invertible since the
arrows \Qn , \Gn\ in diagram \req{diab0ch} cannot be reversed.

One finds similar results associated to the charge $q=1$ singular
vector $\kc^{(1)G}_{l,{\kp^G}}{\ }$ in the Verma module
$V(\ket{\D,\htop}^G)$, as shown in diagrams \req{diab1} and
\req{diab1ch}. For $\D+l \neq 0$ the box diagram consists of \svec s
of types $\kc^{(1)G}_{l,{\kp^G}}{\ }$ and $\kc^{(0)Q}_{l,{\kp^G}}{\ }$
at the same level $l$ in the Verma module $V(\ket{\D,\htop}^G)$, and
\svec s of types $\kc^{(-1)Q}_{l,{\kp^Q}}{\ }$ and
$\kc^{(0)G}_{l,{\kp^Q}}{\ }$ also at level $l$ in the
Verma module $V(\ket{\D,-\htop-\ctop/3}^Q)$.

\vskip .15in

\def\btggob  {\mbox{$\kc_{l,\, \ket{\D,\,\htop}^G}^{(1)G} $}}
\def\btqqob  {\mbox{$\kc_{l,\, \ket{\D,-\htop-{\ctop\over3}}^Q}^{(-1)Q} $}}
\def\btqqmb  {\mbox{$\kc_{l,\, \ket{\D,\,\htop}^G}^{(0)Q} $}}
\def\btggpb  {\mbox{$\kc_{l,\, \ket{\D,-\htop-{\ctop\over3}}^Q}^{(0)G} $}}

  \begin{equation} \begin{array}{rcl}
  \btggob &
  \stackrel{\Qz}{\mbox{------}\!\!\!\longrightarrow}
  & \btqqmb \\[3 mm]
   \cA\,\updownarrow\ && \ \updownarrow\, \cA
  \\[3 mm]  \btqqob \! & \stackrel{\Gz}
  {\mbox{------}\!\!\!\longrightarrow} & \! \btggpb
 \end{array} \label{diab1} \end{equation}

\vskip .15in

For the case of zero conformal weight $\D+l=0$ the uncharged \svec s on
the right-hand side become secondary chiral singular vectors:

\vskip .15in

\def\btggobc  {\mbox{$\kc_{l,\, \ket{-l,\,\htop}^G}^{(1)G} $}}
\def\btqqobc  {\mbox{$\kc_{l,\, \ket{-l,-\htop-{\ctop\over3}}^Q}^{(-1)Q}$}}
\def\btqqmbc  {\mbox{$\kc_{l,\, \ket{-l,\,\htop}^G}^{(0)G,Q} $}}
\def\btggpbc  {\mbox{$\kc_{l,\, \ket{-l,-\htop-{\ctop\over3}}^Q}^{(0)G,Q}$}}

  \begin{equation} \begin{array}{rcl}
  \btggobc &
  \stackrel{\Qz}{\mbox{------}\!\!\!\longrightarrow}
  & \btqqmbc \\[3 mm]
   \cA\,\updownarrow\ && \ \updownarrow\, \cA
  \\[3 mm]  \btqqobc \! & \stackrel{\Gz}
  {\mbox{------}\!\!\!\longrightarrow} & \! \btggpbc
 \end{array} \label{diab1ch} \end{equation}

\vskip .15in

The untwisting of the charged \svec\
$\kc^{(1)G}_{l,\,{\ket{\D,\,\htop}^G}}{\ }$, using $T_W^{\pm }$ \req{twa},
produces two charged mirror-symmetric NS \svec s
located in mirror-symmetric Verma modules. Conversely, the twisting
of two charged mirror-symmetric NS \svec s, using $T_W^+$ and $T_W^-$
respectively, produces the same charged $(|q|=1)$ topological \svec ,
as shown in \req{mtq}. For charge $q=1$ one finds the maps:
\BE \kc^{(1)G}_{l,\,{\ket{\D+\htop/2,\,\htop}^G}} =
 T_W^+ {\ } \ket{\chi_{NS}}^{(1)}_{l-1/2,\,{\ket{\D,\,\htop}}}
 = T_W^- {\ }
 \ket{\chi_{NS}}^{(-1)}_{l-1/2,\,{\ket{\D,-\htop}}}\,. \label{mtch+} \EE
\noi
 where we have redefined $\D$ again, for convenience.
 The maps from the NS \svec s to the remaining topological \svec s
 in diagrams \req{diab1} and \req{diab1ch} are given by:

\BE \begin{array}{lcl}
\kc^{(-1)Q}_{l,\,{\ket{\D+\htop/2,-\htop-{\ctop\over3}}^Q}} &=& \cA {\ }
T_W^+ {\ } \ket{\chi_{NS}}^{(1)}_{l-1/2,\,{\ket{\D,\,\htop}}} \\

\kc^{(0)Q}_{l,\,{\ket{\D+\htop/2,\,\htop}^G}} &=& \Qz {\ }
T_W^+ {\ } \ket{\chi_{NS}}^{(1)}_{l-1/2,\,{\ket{\D,\,\htop}}} \\

\kc^{(0)G}_{l,\,{\ket{\D+\htop/2,-\htop-{\ctop\over3}}^Q}} &=& \cA {\ }
\Qz {\ } T_W^+ {\ } \ket{\chi_{NS}}^{(1)}_{l-1/2,\,{\ket{\D,\,\htop}}} \\

\kc^{(0)G,Q}_{l,\,{\ket{-l,\,\htop}^G}} &=& \Qz {\ }
T_W^+ {\ } \ket{\chi_{NS}}^{(1)}_{l-1/2,\,{\ket{-l-\htop/2,\,\htop}}} \\

\kc^{(0)G,Q}_{l,\,{\ket{-l,-\htop-{\ctop\over3}}^Q}} &=& \cA {\ }\Qz {\ }
T_W^+ {\ } \ket{\chi_{NS}}^{(1)}_{l-1/2,\,{\ket{-l-\htop/2,\,\htop}}}
\label{cfch+} \end{array} \EE

\noi
and mirror-symmetric expressions using
$T_W^- \,\ket{\chi_{NS}}^{(-1)}_{l-1/2,\,{\ket{\D,-\htop}}} $. As before,
the last two maps, to chiral \svec s, are not invertible since the
arrows \Qn , \Gn\ in diagram \req{diab1ch} cannot be reversed.

Finally let us take a charge $q=-1$ singular vector
 $\kc^{(-1)G}_{l,{\kp^G}}{\ }$ in the Verma module
 $V(\ket{\D,\htop}^G)$. For $\D+l \neq 0$ the box diagram
 \req{diab2} consists of \svec s of the types
 $\kc^{(-1)G}_{l,{\kp^G}}{\ }$ and $\kc^{(-2)Q}_{l,{\kp^G}}{\ }$, at the
 same level $l$ in the Verma module $V(\ket{\D,\htop}^G)$, and
 singular vectors of the types
$\kc^{(1)Q}_{l,{\kp^Q}}{\ }$ and $\kc^{(2)G}_{l,{\kp^Q}}{\ }$ also at
level $l$ in the Verma module $V(\ket{\D,-\htop-\ctop/3}^Q)$.

\def\bbggob  {\mbox{$\kc_{l,\, \ket{\D,\,\htop}^G}^{(-1)G} $}}
\def\bbqqob
 {\mbox{$\kc_{l,\, \ket{\D,-\htop-{\ctop\over3}}^Q}^{(1)Q} $}}
\def\bbqqmb
 {\mbox{$\kc_{l,\, \ket{\D,\,\htop}^G}^{(-2)Q} $}}
\def\bbggpb
 {\mbox{$\kc_{l,\, \ket{\D,-\htop-{\ctop\over3}}^Q}^{(2)G} $}}

  \begin{equation} \begin{array}{rcl}
  \bbggob &
  \stackrel{\Qz}{\mbox{------}\!\!\!\longrightarrow}
  & \bbqqmb \\[3 mm]
   \cA\,\updownarrow\ && \ \updownarrow\, \cA
  \\[3 mm]  \bbqqob \! & \stackrel{\Gz}
  {\mbox{------}\!\!\!\longrightarrow} & \! \bbggpb
 \end{array} \label{diab2} \end{equation}

\vskip .15in

In this case the twistings of the two charged mirror-symmetric NS \svec s
which produce the $q=-1$ charged topological \svec\ read:
\BE \kc^{(-1)G}_{l,\,{\ket{\D+\htop/2,\,\htop}^G}} =
 T_W^+ {\ } \ket{\chi_{NS}}^{(-1)}_{l+1/2,\,{\ket{\D,\,\htop}}}
 = T_W^- {\ }
 \ket{\chi_{NS}}^{(1)}_{l+1/2,\,{\ket{\D,-\htop}}}\,.\label{mtch-} \EE

 The maps from the NS \svec s to the remaining topological \svec s
 in diagram \req{diab2} result as follows:

 \BE \begin{array}{lcl}
\kc^{(1)Q}_{l,\,{\ket{\D+\htop/2,-\htop-{\ctop\over3}}^Q}} &=& \cA {\ }
T_W^+ {\ } \ket{\chi_{NS}}^{(-1)}_{l+1/2,\,{\ket{\D,\,\htop}}} \\

\kc^{(-2)Q}_{l,\,{\ket{\D+\htop/2,\,\htop}^G}} &=& \Qz {\ }
T_W^+ {\ } \ket{\chi_{NS}}^{(-1)}_{l+1/2,\,{\ket{\D,\,\htop}}} \\

\kc^{(2)G}_{l,\,{\ket{\D+\htop/2,-\htop-{\ctop\over3}}^Q}} &=& \cA {\ }
  \Qz {\ } T_W^+ {\ } \ket{\chi_{NS}}^{(-1)}_{l+1/2,\,{\ket{\D,\,\htop}}}
\label{cfch-} \end{array} \EE
\noi
and mirror-symmetric expressions using
$T_W^- \,\ket{\chi_{NS}}^{(1)}_{l+1/2,\,{\ket{\D,-\htop}}} $.

Let us come back to diagram \req{diab2}.
For $\D+l=0$ the `would be' secondary chiral singular vectors with $|q|=2$
simply do not exist, as follows from the results in tables \req{tabl2}
and \req{tabl3}. As a consequence, the singular vectors of types
$\kc^{(-1)G}_{l,{\kp^G}}{\ }$ and $\kc^{(1)Q}_{l,{\kp^Q}}{\ }$
`become' actually chiral for zero conformal weight, \ie\ of types
$\kc^{(-1)G,Q}_{l,{\kp^G}}{\ }$ and $\kc^{(1)G,Q}_{l,{\kp^Q}}{\ }$
instead, and the box diagram reduces to two chiral singular vectors,
connected by $\cA$. Thus one has the maps:

\BE \begin{array}{lclcl} \kc^{(-1)G,Q}_{l,\,{\ket{-l,\,\htop}^G}} &=&
 T_W^+ {\ } \ket{\chi_{NS}}^{(-1)}_{l+1/2,\,{\ket{-l-\htop/2,\,\htop}}}
 &=& T_W^- {\ }
\ket{\chi_{NS}}^{(1)}_{l+1/2,\,{\ket{-l-\htop/2,-\htop}}}\, \\
\kc^{(1)G,Q}_{l,\,{\ket{-l,-\htop-{\ctop\over3}}^Q}} &=& \cA {\ }
 T_W^+ {\ } \ket{\chi_{NS}}^{(-1)}_{l+1/2,\,{\ket{-l-\htop/2,\,\htop}}}
 &=& \cA {\ } T_W^- {\ }
\ket{\chi_{NS}}^{(1)}_{l+1/2,\,{\ket{-l-\htop/2,-\htop}}}\,.\label{mtchc}
\end{array} \EE

An important observation now is the following.
The fact that $\kc^{(-1)G}_{l,\,{\ket{\D,\,\htop}^G}}{\ }$ becomes chiral
for $\D=-l$ implies necessarily that all the NS \svec s of the type
$\ket{\chi_{NS}}^{(-1)}_{l+1/2,\,{\ket{-l-\htop/2,\,\htop}}}$ are antichiral
(annihilated by $G^-_{-1/2}$), whereas all the NS \svec s of the type
$\ket{\chi_{NS}}^{(1)}_{l+1/2,\,{\ket{-l-\htop/2,-\htop}}}$
are chiral (annihilated by $G^+_{-1/2}$). The reason is that
$Q_0 = T_W^+\,G^-_{-1/2} = T_W^-\,G^+_{-1/2}$, so that the condition
of being annihilated by \Qn\ is transformed into the conditions of being
annihilated by $G^-_{-1/2}$ and $G^+_{-1/2}$, respectively, under the
twists $T_W^+$ and $T_W^-$. Thus we have found that all the charged NS
\svec s $\ket{\chi_{NS}}^{(\pm 1)}_{l',\,{\ket{\D',\,\htop'}}}$ with
$\D'+l'= \pm {(\htop'\pm 1)\over2}$ are chiral (upper signs) or antichiral
(lower signs). Furthermore, the uncharged chiral singular vectors
are equivalent to charged chiral
singular vectors, as was pointed out in section 2. Namely
\BE \kc^{(0)G,Q}_{l,\,{\ket{-l,\,\htop-1}^Q}} =
\kc^{(-1)G,Q}_{l,\,{\ket{-l,\,\htop}^G}}\,, {\ } \qquad
\kc^{(0)G,Q}_{l,\,{\ket{-l,-\htop-{\ctop\over3}+1}^G}}=
\kc^{(1)G,Q}_{l,\,{\ket{-l,-\htop-{\ctop\over3}}^Q}}\,, \label{mtchu} \EE
\noi
by exchanging the primary states of the Verma module:
$\ket{-l,\,\htop}^G = G_0 \, \ket{-l,\,\htop-1}^Q$ and
$\ket{-l,-\htop-{\ctop\over3}}^Q = Q_0 \,
\ket{-l,-\htop-{\ctop\over3}+1}^G$. As a consequence we have found
invertible maps from the NS \svec s to the chiral topological
\svec s (charged as well as uncharged). These are, in addition, simpler
than the maps \req{cfu} and \req{cfch+} deduced from diagrams
\req{diab0ch} and \req{diab1ch}.

\subsection{Maps from NS to R \svec s}\lvm

The maps from the NS to the R \svec s are derived using the spectral
flows $\cU_{\pm 1/2}$ and $\cA_{\pm 1/2}$, as deduced from
expressions \req{spfl} and \req{ospfl}. To be precise, the R
\svec s with the same helicities than the primaries on which they
are built are directly connected to the NS \svec s through the
spectral flows, whereas the \svec s with different helicities than
their primaries are related to the NS \svec s through the spectral
flows plus one the modes $G_0^+$ or $G_0^-$. Similarly as for the
topological \svec s, in every case there are two mirror-symmetric
NS \svec s mapped to each R \svec .

Let us start with the \svec s of table \req{tabl5}
built on helicity $(+)$ primaries $\ket{\D,\htop}^+$.
The helicity $(+)$ R \svec s $\kcr_l^{(-1)+}$, $\kcr_l^{(0)+}$ and
$\kcr_l^{(1)+}$ are obtained from the  NS \svec s through the maps:
\BE
\kcr^{(q)+}_{l,\,{\ket{\D+{\htop\over2}+
{\ctop\over24},\,\htop+{\ctop\over6}}^+}} =  \cU_{- 1/2} {\ }
\ket{\chi_{NS}}^{(q)}_{l-{q\over2},\,{\ket{\D,\,\htop}}}
 = \cA_{- 1/2} {\ }
 \ket{\chi_{NS}}^{(-q)}_{l-{q\over2},\,{\ket{\D,-\htop}}}\,, \label{aa1} \EE
\noi
resulting in:
\BE \begin{array}{lcl}
{\ }\kcr^{(-1)+}_{l,\,{\ket{\D+{\htop\over2}+
{\ctop\over24},\,\htop+{\ctop\over6}}^+}} &=& \cU_{- 1/2}
{\ } \ket{\chi_{NS}}^{(-1)}_{l+{1\over2},\,{\ket{\D,\,\htop}}} \\

{\ }\kcr^{(0)+}_{l,\,{\ket{\D+{\htop\over2}+
{\ctop\over24},\,\htop+{\ctop\over6}}^+}} &=&
 \cU_{- 1/2} {\ } \ket{\chi_{NS}}^{(0)}_{l,\,{\ket{\D,\,\htop}}} \\

{\ }\kcr^{(1)+}_{l,\,{\ket{\D+{\htop\over2}+
{\ctop\over24},\,\htop+{\ctop\over6}}^+}} &=&
 \cU_{- 1/2} {\ } \ket{\chi_{NS}}^{(1)}_{l-{1\over2},\,{\ket{\D,\,\htop}}}
\label{aa2} \end{array} \EE

\noi
and mirror-symmetric expressions using
$\cA_{-1/2} \, \ket{\chi_{NS}}^{(-q)}_{l-{q\over2},\,{\ket{\D,-\htop}}}$.
The helicity $(-)$ R \svec s $\kcr_l^{(-2)-}$, $\kcr_l^{(-1)-}$ and
$\kcr_l^{(0)-}$ are obtained from the previous by the action of $G_0^-$
since
\BE
\kcr^{(q-1)-}_{l,\,{\ket{\D+{\htop\over2}+
{\ctop\over24},\,\htop+{\ctop\over6}}^+}} = \ G_0^- \
\kcr^{(q)+}_{l,\,{\ket{\D+{\htop\over2}+
{\ctop\over24},\,\htop+{\ctop\over6}}^+}} =  \ G_0^- \
 \cU_{- 1/2} {\ } \ket{\chi_{NS}}^{(q)}_{l-{q\over2},\,
{\ket{\D,\,\htop}}} \,,\label{aa3} \EE
\noi
that is:
\BE \begin{array}{lcl}
{\ }\kcr^{(-2)-}_{l,\,{\ket{\D+{\htop\over2}+
{\ctop\over24},\,\htop+{\ctop\over6}}^+}} &=& \ G_0^- \ \cU_{- 1/2} {\ }
\ket{\chi_{NS}}^{(-1)}_{l+{1\over2},\,{\ket{\D,\,\htop}}} \\

{\ }\kcr^{(-1)-}_{l,\,{\ket{\D+{\htop\over2}+
{\ctop\over24},\,\htop+{\ctop\over6}}^+}} &=& \ G_0^- \
 \cU_{- 1/2} {\ } \ket{\chi_{NS}}^{(0)}_{l,\,{\ket{\D,\,\htop}}} \\

{\ }\kcr^{(0)-}_{l,\,{\ket{\D+{\htop\over2}+
{\ctop\over24},\,\htop+{\ctop\over6}}^+}} &=& \ G_0^- \  \cU_{- 1/2}
{\ } \ket{\chi_{NS}}^{(1)}_{l-{1\over2},\,{\ket{\D,\,\htop}}}
\label{aa4} \end{array} \EE

\noi
and mirror-symmetric expressions using
$\ G_0^- \cA_{-1/2} \,
\ket{\chi_{NS}}^{(-q)}_{l-{q\over2},\,{\ket{\D,-\htop}}}$.
When the total conformal weight of the \svec s of types $\kcr^{(0)-}_l$
and $\kcr^{(-1)+}_l$ is equal to ${\ctop\over24}$ then they `become' the
chiral \svec s $\kcr^{(0)+,-}_l$ and $\kcr^{(-1)+,-}_l$.
As a result one gets the maps:
\BE \begin{array}{lcl}
{\ }\kcr^{(0)+,-}_{l,\,{\ket{{\ctop\over24}-l,
\,\htop+{\ctop\over6}}^+}} &=& \ G_0^- \  \cU_{- 1/2}
{\ } \ket{\chi_{NS}}^{(1)}_{l-{1\over2},\,{\ket{-{\htop\over2}-l,\,\htop}}} \\ 
{\ }\kcr^{(-1)+,-}_{l,\,{\ket{{\ctop\over24}-l,
\,\htop+{\ctop\over6}}^+}} &=& \cU_{- 1/2}
{\ } \ket{\chi_{NS}}^{(-1)}_{l+{1\over2},\,{\ket{-{\htop\over2}-l,\,\htop}}}   
\label{aa5} \end{array} \EE

\noi
and mirror-symmetric expressions using
$\cA_{-1/2} \,
\ket{\chi_{NS}}^{(-q)}_{l-{q\over2},\,{\ket{-{\htop\over2}-l,-\htop}}}$.

Observe that the map to the charged chiral \svec\ $\kcr^{(-1)+,-}_l$
is invertible while the map to the uncharged $\kcr^{(0)+,-}_l$ is not
invertible as the action of $G_0^-$ cannot be reversed in this case
(since chiral \svec s are annihilated by $G_0^+$ as well as by $G_0^-$).
Nevertheless, similarly as we showed in the topological case, the
uncharged chiral \svec s can be `converted' into charged chiral \svec s.
We will come back to this issue a few paragraphs below.

Now let us move to the \svec s of table \req{tabl6}
built on helicity $(-)$ primaries $\ket{\D,\htop}^-$.
The helicity $(-)$ R \svec s $\kcr_l^{(-1)-}$, $\kcr_l^{(0)-}$ and
$\kcr_l^{(1)-}$ are obtained from the  NS \svec s through the maps:
\BE
\kcr^{(q)-}_{l,\,{\ket{\D-{\htop\over2}+
{\ctop\over24},\,\htop-{\ctop\over6}}^-}} =  \cU_{1/2} {\ }
\ket{\chi_{NS}}^{(q)}_{l+{q\over2},\,{\ket{\D,\,\htop}}}
 = \cA_{1/2} {\ }
 \ket{\chi_{NS}}^{(-q)}_{l+{q\over2},\,{\ket{\D,-\htop}}}\,, \label{bb1} \EE
\noi
which give:
\BE \begin{array}{lcl}
{\ }\kcr^{(-1)-}_{l,\,{\ket{\D-{\htop\over2}+
{\ctop\over24},\,\htop-{\ctop\over6}}^-}} &=& \cU_{1/2}
{\ } \ket{\chi_{NS}}^{(-1)}_{l-{1\over2},\,{\ket{\D,\,\htop}}} \\

{\ }\kcr^{(0)-}_{l,\,{\ket{\D-{\htop\over2}+
{\ctop\over24},\,\htop-{\ctop\over6}}^-}} &=&
 \cU_{1/2} {\ } \ket{\chi_{NS}}^{(0)}_{l,\,{\ket{\D,\,\htop}}} \\

{\ }\kcr^{(1)-}_{l,\,{\ket{\D-{\htop\over2}+
{\ctop\over24},\,\htop-{\ctop\over6}}^-}} &=&
 \cU_{1/2} {\ } \ket{\chi_{NS}}^{(1)}_{l+{1\over2},\,{\ket{\D,\,\htop}}}
\label{bb2} \end{array} \EE

\noi
and mirror-symmetric expressions using $\cA_{1/2} \,
\ket{\chi_{NS}}^{(-q)}_{l+{q\over2},\,{\ket{\D,-\htop}}}$.
The helicity $(+)$ R \svec s $\kcr_l^{(0)+}$, $\kcr_l^{(1)+}$ and
$\kcr_l^{(2)+}$ are obtained now simply by acting with $G_0^+$:
\BE
\kcr^{(q+1)+}_{l,\,{\ket{\D-{\htop\over2}+
{\ctop\over24},\,\htop-{\ctop\over6}}^-}} = \ G_0^+ \
\kcr^{(q)-}_{l,\,{\ket{\D-{\htop\over2}+
{\ctop\over24},\,\htop-{\ctop\over6}}^-}} =  \ G_0^+ \
 \cU_{1/2} {\ } \ket{\chi_{NS}}^{(q)}_{l+{q\over2},\,{\ket{\D,\,\htop}}} \,,
 \label{bb3} \EE
\noi
resulting in:
 \BE \begin{array}{lcl}
{\ }\kcr^{(0)+}_{l,\,{\ket{\D-{\htop\over2}+
{\ctop\over24},\,\htop-{\ctop\over6}}^-}} &=& G_0^+ \cU_{1/2}
{\ } \ket{\chi_{NS}}^{(-1)}_{l-{1\over2},\,{\ket{\D,\,\htop}}} \\

{\ }\kcr^{(1)+}_{l,\,{\ket{\D-{\htop\over2}+
{\ctop\over24},\,\htop-{\ctop\over6}}^-}} &=& G_0^+
 \cU_{1/2} {\ } \ket{\chi_{NS}}^{(0)}_{l,\,{\ket{\D,\,\htop}}} \\

{\ }\kcr^{(2)+}_{l,\,{\ket{\D-{\htop\over2}+
{\ctop\over24},\,\htop-{\ctop\over6}}^-}} &=& G_0^+
 \cU_{1/2} {\ } \ket{\chi_{NS}}^{(1)}_{l+{1\over2},\,{\ket{\D,\,\htop}}}
\label{bb4} \end{array} \EE

\noi
and mirror-symmetric expressions using $\ G_0^+\cA_{1/2} \,
\ket{\chi_{NS}}^{(-q)}_{l+{q\over2},\,{\ket{\D,-\htop}}}$.
When the total conformal weight of the \svec s of types $\kcr^{(0)+}_l$
and $\kcr^{(1)-}_l$ is equal to ${\ctop\over24}$ then they `become' the
chiral \svec s $\kcr^{(0)+,-}_l$ and $\kcr^{(1)+,-}_l$.
As a result one gets the maps:
\BE \begin{array}{lcl}
{\ }\kcr^{(0)+,-}_{l,\,{\ket{{\ctop\over24}-l,
\,\htop-{\ctop\over6}}^-}} &=& G_0^+ \cU_{1/2}
{\ } \ket{\chi_{NS}}^{(-1)}_{l-{1\over2},\,{\ket{{\htop\over2}-l,\,\htop}}} \\  

{\ }\kcr^{(1)+,-}_{l,\,{\ket{{\ctop\over24}-l,
\,\htop-{\ctop\over6}}^-}} &=&  \cU_{1/2} {\ }
\ket{\chi_{NS}}^{(1)}_{l+{1\over2},\,{\ket{{\htop\over2}-l,\,\htop}}}

\label{bb5} \end{array} \EE

\noi
and mirror-symmetric expressions using
$\cA_{1/2} \,
\ket{\chi_{NS}}^{(-q)}_{l+{q\over2},\,{\ket{{\htop\over2}-l,-\htop}}}$.

Observe that, again, the map to the charged chiral \svec\
$\kcr^{(1)+,-}_l$ is invertible while the map to the uncharged
$\kcr^{(0)+,-}_l$ is not. Nevertheless the uncharged chiral \svec s
$\kcr^{(0)+,-}_l$, from tables \req{tabl5} and \req{tabl6}, are equivalent
to charged chiral \svec s as there are always two h.w. vectors with
opposite helicities in the Verma modules which contain chiral \svec s.
Thus one only has to express the uncharged \svec\ as built on the
h.w. vector with the opposite helicity than previously and it will
pick up a nonzero $q$. In this manner the \svec s $\kcr^{(0)+,-}_l$ on
helicity $(-)$ primaries are equivalent to \svec s $\kcr^{(-1)+,-}_l$
on helicity $(+)$ primaries and the \svec s $\kcr^{(0)+,-}_l$ on
helicity $(+)$ primaries are equivalent to \svec s $\kcr^{(1)+,-}_l$
on helicity $(-)$ primaries. Thus one has the invertible maps:

\BE \begin{array}{lclcl}
{\ }\kcr^{(0)+,-}_{l,\,{\ket{{\ctop\over24}-l,
\,\htop-{\ctop\over6}+1}^+}} &=&
{\ }\kcr^{(1)+,-}_{l,\,{\ket{{\ctop\over24}-l,
\,\htop-{\ctop\over6}}^-}} &=&  \cU_{1/2} {\ }
\ket{\chi_{NS}}^{(1)}_{l+{1\over2},\,{\ket{{\htop\over2}-l,\,\htop}}} \\

{\ }\kcr^{(0)+,-}_{l,\,{\ket{{\ctop\over24}-l,
\,\htop+{\ctop\over6}-1}^-}} &=&
{\ }\kcr^{(-1)+,-}_{l,\,{\ket{{\ctop\over24}-l,
\,\htop+{\ctop\over6}}^+}} &=& \cU_{- 1/2} {\ }
\ket{\chi_{NS}}^{(-1)}_{l+{1\over2},\,
{\ket{-{\htop\over2}-l,\,\htop}}}
\label{cc1} \end{array} \EE
\noi
and mirror-symmetric expressions using $\cA_{1/2}$ and $\cA_{-1/2}$,
respectively.

\subsection{Construction Formulae}\lvm

The maps given in subsections 3.1 and 3.2 turn into construction
formulae for the $16+16$ types of \svec s of the Topological
and of the Ramond N=2 algebras just by expressing the
singular vectors of the Neveu-Schwarz N=2 algebra in terms of their
corresponding construction formulae. In what follows we will
describe briefly the construction formulae for the NS \svec s --
all the details are in refs. \cite{Doerr1} and \cite{Doerr2} --
and we will summarize the maps that provide construction formulae
for the topological and for the R \svec s, in each case.

\subsubsection{Charged NS \svec s $\ket{\chi_{NS}}^{(\pm1)}$}

The explicit expressions for the charged \svec s at level $k$ read
\cite{Doerr1}
\BE \ket{\chi_{NS}}^{(\pm1)}_k= {\cal W}^{\pm} {\cal E}^{\pm}(k-1/2)
{\cal T}^{\pm}(k-1){\cal E}^{\pm}(k-3/2){\cal T}^{\pm}(k-2) ....
{\cal E}^{\pm}(1) {\cal T}^{\pm}(1/2) \Psi_0^{\pm} \,, \label{dd1}\EE

\noi
where ${\cal E}^{\pm}(k)$ and ${\cal T}^{\pm}(k)$
are even and odd recursion step matrices, respectively,
and ${\cal W}^{\pm}$ and $\Psi_0^{\pm}$ are
vectors, the latter depending on the initial low level \svec s. The
spectrum of $\D$ and $\htop$ for which the NS Verma modules
$V_{NS}(\D,\,\htop)$ contain charged \svec s $\ket{\chi_{NS}}^{(\pm1)}_k$
is given (at least) by the zeroes of the NS determinant formula which are
solutions to the vanishing planes $g_{\pm k}(\D,\,\htop)=0$
\cite{BFK}\cite{Nam}.

As follows from the results of subsections 3.1 and 3.2, the construction
formulae \req{dd1} for charged NS \svec s provide also construction
formulae for the following topological \svec s:

\BE \begin{array}{lclcl}

\kc^{(1)G}_{l,\,{\ket{\D+\htop/2,\,\htop}^G}} &=&
 T_W^+  \ket{\chi_{NS}}^{(1)}_{l-1/2,\,{\ket{\D,\,\htop}}} &=&
 T_W^-  \ket{\chi_{NS}}^{(-1)}_{l-1/2,\,{\ket{\D,-\htop}}} \\

\kc^{(-1)Q}_{l,\,{\ket{\D+\htop/2,-\htop-{\ctop\over3}}^Q}} &=& \cA
T_W^+  \ket{\chi_{NS}}^{(1)}_{l-1/2,\,{\ket{\D,\,\htop}}} &=& \cA
T_W^-  \ket{\chi_{NS}}^{(-1)}_{l-1/2,\,{\ket{\D,\,-\htop}}}\\

\kc^{(0)Q}_{l,\,{\ket{\D+\htop/2,\,\htop}^G}} &=& \Qz
T_W^+  \ket{\chi_{NS}}^{(1)}_{l-1/2,\,{\ket{\D,\,\htop}}} &=& \Qz
T_W^-  \ket{\chi_{NS}}^{(-1)}_{l-1/2,\,{\ket{\D,\,-\htop}}}\\

\kc^{(0)G}_{l,\,{\ket{\D+\htop/2,-\htop-{\ctop\over3}}^Q}} &=& \cA
\Qz  T_W^+  \ket{\chi_{NS}}^{(1)}_{l-1/2,\,{\ket{\D,\,\htop}}}
&=& \cA  \Qz
T_W^-  \ket{\chi_{NS}}^{(-1)}_{l-1/2,\,{\ket{\D,\,-\htop}}}

\end{array} \EE

 \BE \begin{array}{lclcl}

 \kc^{(-1)G}_{l,\,{\ket{\D+\htop/2,\,\htop}^G}} &=&
 T_W^+  \ket{\chi_{NS}}^{(-1)}_{l+1/2,\,{\ket{\D,\,\htop}}}
 &=& T_W^-  \ket{\chi_{NS}}^{(1)}_{l+1/2,\,{\ket{\D,-\htop}}} \\

\kc^{(1)Q}_{l,\,{\ket{\D+\htop/2,-\htop-{\ctop\over3}}^Q}} &=& \cA
T_W^+  \ket{\chi_{NS}}^{(-1)}_{l+1/2,\,{\ket{\D,\,\htop}}} &=& \cA
T_W^-  \ket{\chi_{NS}}^{(1)}_{l+1/2,\,{\ket{\D,-\htop}}} \\

\kc^{(-2)Q}_{l,\,{\ket{\D+\htop/2,\,\htop}^G}} &=& \Qz
T_W^+  \ket{\chi_{NS}}^{(-1)}_{l+1/2,\,{\ket{\D,\,\htop}}} &=& \Qz
T_W^-  \ket{\chi_{NS}}^{(1)}_{l+1/2,\,{\ket{\D,-\htop}}} \\

\kc^{(2)G}_{l,\,{\ket{\D+\htop/2,-\htop-{\ctop\over3}}^Q}} &=& \cA  \Qz
T_W^+  \ket{\chi_{NS}}^{(-1)}_{l+1/2,\,{\ket{\D,\,\htop}}}
&=& \cA \Qz T_W^- \ket{\chi_{NS}}^{(1)}_{l+1/2,\,{\ket{\D,-\htop}}} \\

\kc^{(-1)G,Q}_{l,\,{\ket{-l,\,\htop}^G}} &=& T_W^+
\ket{\chi_{NS}}^{(-1)}_{l+1/2,\,{\ket{-l-\htop/2,\,\htop}}} &=& T_W^-
\ket{\chi_{NS}}^{(1)}_{l+1/2,\,{\ket{-l-\htop/2,-\htop}}}\, \\

\kc^{(1)G,Q}_{l,\,{\ket{-l,-\htop-{\ctop\over3}}^Q}} &=& \cA
 T_W^+  \ket{\chi_{NS}}^{(-1)}_{l+1/2,\,{\ket{-l-\htop/2,\,\htop}}}
 &=& \cA T_W^-
\ket{\chi_{NS}}^{(1)}_{l+1/2,\,{\ket{-l-\htop/2,-\htop}}} \\

 \kc^{(0)G,Q}_{l,\,{\ket{-l,\,\htop-1}^Q}} &=& {\ }
\kc^{(-1)G,Q}_{l,\,{\ket{-l,\,\htop}^G}} & {\ } & {\ } \\

\kc^{(0)G,Q}_{l,\,{\ket{-l,-\htop-{\ctop\over3}+1}^G}} &=& {\ }
\kc^{(1)G,Q}_{l,\,{\ket{-l,-\htop-{\ctop\over3}}^Q}}\ , & {\ } & {\ }

\end{array} \EE

\noi
and for the following R \svec s:

\BE \begin{array}{lclcl}

\kcr^{(1)+}_{l,\,{\ket{\D+{\htop\over2}+
{\ctop\over24},\,\htop+{\ctop\over6}}^+}} &=& \cU_{- 1/2}
 \ket{\chi_{NS}}^{(1)}_{l-{1\over2},\,{\ket{\D,\,\htop}}} &=&
\cA_{-1/2} \ket{\chi_{NS}}^{(-1)}_{l-{1\over2},\,{\ket{\D,-\htop}}} \\

\kcr^{(-1)+}_{l,\,{\ket{\D+{\htop\over2}+
{\ctop\over24},\,\htop+{\ctop\over6}}^+}} &=& \cU_{- 1/2}
\ket{\chi_{NS}}^{(-1)}_{l+{1\over2},\,{\ket{\D,\,\htop}}} &=&
\cA_{-1/2} \ket{\chi_{NS}}^{(1)}_{l+{1\over2},\,{\ket{\D,-\htop}}} \\

\kcr^{(0)-}_{l,\,{\ket{\D+{\htop\over2}+
{\ctop\over24},\,\htop+{\ctop\over6}}^+}} &=& G_0^- \cU_{- 1/2}
\ket{\chi_{NS}}^{(1)}_{l-{1\over2},\,{\ket{\D,\,\htop}}}
&=&  G_0^- \cA_{-1/2}
\ket{\chi_{NS}}^{(-1)}_{l-{1\over2},\,{\ket{\D,-\htop}}} \\

\kcr^{(-2)-}_{l,\,{\ket{\D+{\htop\over2}+
{\ctop\over24},\,\htop+{\ctop\over6}}^+}} &=&  G_0^- \cU_{- 1/2}
\ket{\chi_{NS}}^{(-1)}_{l+{1\over2},\,{\ket{\D,\,\htop}}}
&=&  G_0^-
\cA_{-1/2} \ket{\chi_{NS}}^{(1)}_{l+{1\over2},\,{\ket{\D,-\htop}}} \\

\kcr^{(-1)-}_{l,\,{\ket{\D-{\htop\over2}+
{\ctop\over24},\,\htop-{\ctop\over6}}^-}} &=& \cU_{1/2}
\ket{\chi_{NS}}^{(-1)}_{l-{1\over2},\,{\ket{\D,\,\htop}}} &=&
\cA_{1/2} \ket{\chi_{NS}}^{(1)}_{l-{1\over2},\,{\ket{\D,\,-\htop}}} \\

\kcr^{(1)-}_{l,\,{\ket{\D-{\htop\over2}+
{\ctop\over24},\,\htop-{\ctop\over6}}^-}} &=&  \cU_{1/2}
\ket{\chi_{NS}}^{(1)}_{l+{1\over2},\,{\ket{\D,\,\htop}}} &=& \cA_{1/2}
\ket{\chi_{NS}}^{(-1)}_{l+{1\over2},\,{\ket{\D,\,-\htop}}} \\

\kcr^{(0)+}_{l,\,{\ket{\D-{\htop\over2}+
{\ctop\over24},\,\htop-{\ctop\over6}}^-}} &=& G_0^+ \cU_{1/2}
 \ket{\chi_{NS}}^{(-1)}_{l-{1\over2},\,{\ket{\D,\,\htop}}} &=& G_0^+
\cA_{1/2} \ket{\chi_{NS}}^{(1)}_{l-{1\over2},\,{\ket{\D,\,-\htop}}} \\

\kcr^{(2)+}_{l,\,{\ket{\D-{\htop\over2}+
{\ctop\over24},\,\htop-{\ctop\over6}}^-}} &=& G_0^+ \cU_{1/2}
\ket{\chi_{NS}}^{(1)}_{l+{1\over2},\,{\ket{\D,\,\htop}}} &=& G_0^+
\cA_{1/2} \ket{\chi_{NS}}^{(-1)}_{l+{1\over2},\,{\ket{\D,\,-\htop}}} \\

\kcr^{(1)+,-}_{l,\,{\ket{{\ctop\over24}-l,
\,\htop-{\ctop\over6}}^-}} &=& \cU_{1/2}
\ket{\chi_{NS}}^{(1)}_{l+{1\over2},\,{\ket{{\htop\over2}-l,\,\htop}}}
 &=& \cA_{1/2}
\ket{\chi_{NS}}^{(-1)}_{l+{1\over2},\,{\ket{{\htop\over2}-l,\,-\htop}}} \\

\kcr^{(-1)+,-}_{l,\,{\ket{{\ctop\over24}-l,
\,\htop+{\ctop\over6}}^+}} &=& \cU_{- 1/2}
\ket{\chi_{NS}}^{(-1)}_{l+{1\over2},\,{\ket{-{\htop\over2}-l,\,\htop}}}
 &=& \cA_{1/2}
\ket{\chi_{NS}}^{(1)}_{l+{1\over2},\,{\ket{-{\htop\over2}-l,\,-\htop}}} \\

\kcr^{(0)+,-}_{l,\,{\ket{{\ctop\over24}-l,
\,\htop-{\ctop\over6}+1}^+}} &=& {\ }
\kcr^{(1)+,-}_{l,\,{\ket{{\ctop\over24}-l,
\,\htop-{\ctop\over6}}^-}} &{\ } & {\ } \\

\kcr^{(0)+,-}_{l,\,{\ket{{\ctop\over24}-l,
\,\htop+{\ctop\over6}-1}^-}} &=& {\ }
\kcr^{(-1)+,-}_{l,\,{\ket{{\ctop\over24}-l,
\,\htop+{\ctop\over6}}^+}} & {\ } & {\ }.

\end{array} \EE

\subsubsection{Uncharged NS \svec s $\ket{\chi_{NS}}^{(0)}$}

The explicit expressions for the uncharged \svec s at level
$l={rs\over2}$ read \cite{Doerr2}
\BE \ket{\chi_{NS}}^{(0)}_{r,s}=
\epsilon_{r,s}^{+}(t,\,\htop) {\ } \D_{r,s}(1,0){\ } + {\ }
\epsilon_{r,s}^{-}(t,\,\htop) {\ } \D_{r,s}(0,1)\,, \label{dd2}
\EE

\noi
where $\D_{r,s}(1,0)$ and $\D_{r,s}(0,1)$ are two basis vectors taken from
the analytically continued Verma module,
$t={3-\ctop\over3}$ parametrizes the central charge, and
\BE
\epsilon_{r,s}^{\pm}(t,\htop)= \prod_{m=1}^r
(\pm{s-rt\over2t}+{\htop\over t}\mp{1\over2}\pm m)\,\  ,\ r\in\oZ^+,
\,\, s\in2\oZ^+  \,.
\label{Dcond}
\EE

The spectrum of $\D$ and $\htop$ for which the NS Verma modules
$V_{NS}(\D,\,\htop)$ contain uncharged \svec s
$\ket{\chi_{NS}}^{(0)}_{r,s}$ is given (at least) by the zeroes of the NS
determinant formula which are solutions to the quadratic vanishing
surface $f_{r,s}(\D,\,\htop)=0$ \cite{BFK}\cite{Nam}.
Interestingly enough, the simultaneous vanishing of the two curves,
$\epsilon_{r,s}^+(t,\htop)=0$ and $\epsilon_{r,s}^-(t,\htop)=0$,
leads to the appearance of two linearly independent uncharged NS
singular vectors at the same level, in the same Verma module \cite{Doerr2}.
The topological twists $T_W^{\pm }$ \req{twa} let
these conditions invariant, extending the existence of the two-dimensional
space of singular vectors to the topological singular vectors of types
$\kc_{\kp^G}^{(0)G}$, $\kc_{\kp^G}^{(-1)Q}$, $\kc_{\kp^Q}^{(1)G}$ and
$\kc_{\kp^Q}^{(0)Q} \,$, as the generic uncharged NS \svec s are transformed
necessarily into these four types of topological \svec s via the
mappings \req{mtu} and \req{cfu}. Similarly, for the case of the R
\svec s the two-dimensional singular spaces exist for \svec s of
the types $\kcr_{\kp^+}^{(0)+}$, $\kcr_{\kp^+}^{(-1)-}$,
$\kcr_{\kp^-}^{(1)+}$ and $\kcr_{\kp^-}^{(0)-} \,$.

As follows from the results of subsections 3.1 and 3.2, the construction
formulae \req{dd2} for uncharged NS \svec s provide also construction
formulae for the following topological \svec s:

\BE \begin{array}{lclcl}

\kc^{(0)G}_{l,\,{\ket{\D+\htop/2,\,\htop}^G}} &=&
 T_W^+  \ket{\chi_{NS}}^{(0)}_{l,\,{\ket{\D,\,\htop}}}
 &=& T_W^-  \ket{\chi_{NS}}^{(0)}_{l,\,{\ket{\D,-\htop}}} \\

\kc^{(0)Q}_{l,\,{\ket{\D+\htop/2,-\htop-{\ctop\over3}}^Q}} &=&
\cA  T_W^+  \ket{\chi_{NS}}^{(0)}_{l,\,{\ket{\D,\,\htop}}}
&=& \cA  T_W^-  \ket{\chi_{NS}}^{(0)}_{l,\,{\ket{\D,-\htop}}} \\

 \kc^{(-1)Q}_{l,\,{\ket{\D+\htop/2,\,\htop}^G}} &=&
\Qz  T_W^+  \ket{\chi_{NS}}^{(0)}_{l,\,{\ket{\D,\,\htop}}}
&=& \Qz  T_W^-  \ket{\chi_{NS}}^{(0)}_{l,\,{\ket{\D,-\htop}}} \\

\kc^{(1)G}_{l,\,{\ket{\D+\htop/2,-\htop-{\ctop\over3}}^Q}} &=&
\cA \Qz  T_W^+  \ket{\chi_{NS}}^{(0)}_{l,\,{\ket{\D,\,\htop}}}
&=& \cA \Qz T_W^-  \ket{\chi_{NS}}^{(0)}_{l,\,{\ket{\D,-\htop}}} \ ,

\end{array} \EE

\noi
and for the following R \svec s:

\BE \begin{array}{lclcl}

\kcr^{(0)+}_{l,\,{\ket{\D+{\htop\over2}+
{\ctop\over24},\,\htop+{\ctop\over6}}^+}} &=&
 \cU_{- 1/2}  \ket{\chi_{NS}}^{(0)}_{l,\,{\ket{\D,\,\htop}}}  &=&
\cA_{-1/2} \ket{\chi_{NS}}^{(0)}_{l,\,{\ket{\D,-\htop}}} \\

\kcr^{(-1)-}_{l,\,{\ket{\D+{\htop\over2}+
{\ctop\over24},\,\htop+{\ctop\over6}}^+}} &=& G_0^-
 \cU_{- 1/2}  \ket{\chi_{NS}}^{(0)}_{l,\,{\ket{\D,\,\htop}}}  &=&
\ G_0^-  \cA_{-1/2}  \ket{\chi_{NS}}^{(0)}_{l,\,{\ket{\D,-\htop}}} \\

\kcr^{(0)-}_{l,\,{\ket{\D-{\htop\over2}+
{\ctop\over24},\,\htop-{\ctop\over6}}^-}} &=&
 \cU_{1/2}  \ket{\chi_{NS}}^{(0)}_{l,\,{\ket{\D,\,\htop}}}   &=&
 \cA_{1/2}  \ket{\chi_{NS}}^{(0)}_{l,\,{\ket{\D,-\htop}}} \\

\kcr^{(1)+}_{l,\,{\ket{\D-{\htop\over2}+
{\ctop\over24},\,\htop-{\ctop\over6}}^-}} &=& G_0^+
 \cU_{1/2}  \ket{\chi_{NS}}^{(0)}_{l,\,{\ket{\D,\,\htop}}} &=& G_0^+
\cA_{1/2}  \ket{\chi_{NS}}^{(0)}_{l,\,{\ket{\D,\,-\htop}}} \,.

\end{array} \EE

\section{Final Remarks}\lvm

We have written down one-to-one invertible maps between the
\svec s of the Neveu-Schwarz N=2 algebra and $16+16$ types
of singular vectors of the Topological and of the Ramond N=2 algebras.
To be precise, we have derived two mirror-symmetric maps from the
NS \svec s to each of the $16+16$ topological and R \svec s. These
maps provide construction formulae for the latter once the NS \svec s
are expressed in terms of their construction formulae themselves.
The indecomposable no-label and no-helicity \svec s can only be mapped
to NS subsingular vectors \cite{DB1}, for which no construction
formulae exist.

One remarkable finding resulting from our analysis is that the charged
NS \svec s $\ket{\chi_{NS}}^{(\pm 1)}_{l,\,{\ket{\D,\,\htop}}}$ with
$\D+l= \pm {(\htop\pm 1)\over2}$ are {\it necessarily}
chiral (upper signs) or antichiral (lower signs).

\vskip .3in
\centerline{\bf Acknowledgements}

I am grateful to M. D\"orrzapf for reading the preliminar version of
this manuscript and for several suggestions.


\vskip .17in


\begin{thebibliography}{9}
\def\NPB{Nucl. Phys. B}
\def\PLB{Phys. Lett. B}
\def\MPLA{Mod. Phys. Lett. A}
\def\IJMPA{Int. J. Mod. Phys. A}

\bibitem{BPZ} A. Belavin, A. Polyakov and Al. Zamolodchikov, \NPB241
(1984) 333

\bibitem{MM} S. Mukherji, S. Mukhi and A. Sen, \PLB266 (1991) 337

\bibitem{BeSe} B. Gato-Rivera and A.M. Semikhatov, \PLB293 (1992) 72,
Theor. Math. Phys. 95 (1993) 536; \NPB408 (1993) 133

\bibitem{VirSV} L.~Benoit and Y.~Saint-Aubin, \PLB215 (1988) 517;\\
 A.~Kent, \PLB273 (1991) 56;\\
 M.~Bauer, P.~di~Francesco, C.~Itzykson, and J.-B.~Zuber,
\NPB362 (1991) 515

\bibitem{MFF} F.G.~Malikov, B.L.~Feigin, and D.B.~Fuchs, Funk.\ An.\
Prilozh.\ 20 N2 (1986) 25

\bibitem{BaSo} M.~Bauer and N.~Sochen, Comm. Math. Phys. 152 (1993) 127;
\PLB275 (1992) 82

\bibitem{N1S} L. Benoit and Y. Saint-Aubin \IJMPA7 (1992) 3023;
 Lett. Math. Phys. Vol. 23 (1991) 117; \\
 C.S. Huang, D.H. Zhang and Q.R. Zheng, \NPB389 (1993) 81; \\
 G.M.T.~Watts, Nucl. Phys. B407 (1993) 213

\bibitem{Doerr1} M.~D\"orrzapf, \IJMPA10 (1995) 2143

\bibitem{Doerr2} M.~D\"orrzapf, Comm. Math. Phys. 180 (1996) 195

\bibitem{Walg} P.~Bowcock and G.M.T.~Watts, \PLB297 (1992) 282; \\
 Z.~Bajnok, \PLB320 (1994) 36; \PLB329 (1994) 225

\bibitem{KaMa1} M. Kato and S. Matsuda, \PLB172 (1986) 216;
Advanced Studies in Pure Mathematics 16 (1988) 205

\bibitem{GP} A.Ch.~Ganchev and V.B.~Petkova, \PLB293 (1992) 56; \PLB318
(1993) 77

\bibitem{FGP} P.~Furlan, A.Ch.~Ganchev, and V.B.~Petkova,
 \NPB431 (1994) 622-666

\bibitem{Ade} M. Ademollo et al.,\PLB62 (1976) 105, \NPB111 (1976) 77;\\
\NPB114 (1976) 297

\bibitem{Marcus} L. Brink and J.H. Schwarz, \NPB121 (1977) 285;\\
E.S. Fradkin and A.A. Tseytlin, \PLB106 (1981) 63; \PLB 162 (1985) 295;\\
S.D. Mathur and S. Mukhi, Pys. Rev. D 36 (1987) 465; \NPB302 (1988) 130;\\
H. Ooguri and C. Vafa, \NPB361 (1991) 469; \NPB367 (1991) 83;\\
N. Marcus, talk at the Rome String Theory Workshop (1992), hep-th/9211059

\bibitem{Marti} E. Martinec, {\it M-theory and N=2 Strings},
hep-th/9710122 (1997), and references there

\bibitem{HW} C. Hull and E. Witten, \PLB160 (1985) 398

\bibitem{BLNW} M. Bershadsky, W. Lerche, D. Nemeschansky and N.P. Warner,
\NPB401 (1993) 304

\bibitem{BJI6} B. Gato-Rivera and J.I. Rosado, \NPB514 (1998) 477

\bibitem{DB2} M. D\"orrzapf and B. Gato-Rivera, Commun. Math. Phys.
206, (1999) 493

\bibitem{SS} A. Schwimmer and N. Seiberg, \PLB184 (1987) 191

\bibitem{LVW} W.~Lerche, C.~Vafa and N.~P.~Warner,
 \NPB324 (1989) 427

\bibitem{B1} B. Gato-Rivera, \NPB512 (1998) 431

\bibitem{BJI4} B. Gato-Rivera and J.I. Rosado,
 Mod. Phys. Lett. A11 (1996) 423

\bibitem{Kir1} E.B. Kiritsis, Phys. Rev. D36 (1987) 3048

\bibitem{BFK} W. Boucher, D. Friedan and A. Kent, \PLB172 (1986) 316

\bibitem{Nam} S. Nam, \PLB172 (1986) 323

\bibitem{DVV} R. Dijkgraaf, E. Verlinde and H. Verlinde, \NPB352
(1991) 59

\bibitem{DB4} M. D\"orrzapf and B. Gato-Rivera, Comm. Math. Phys.
220 (2001) 263-292

\bibitem{BJI3} B. Gato-Rivera and J.I. Rosado, \PLB369 (1996) 7

\bibitem{BJI5} B. Gato-Rivera and J.I. Rosado, \NPB503 (1997) 447

\bibitem{DB1} M. D\"orrzapf and B. Gato-Rivera, \NPB557 (1999) 517

\bibitem{DB3} M. D\"orrzapf and B. Gato-Rivera, \NPB558 (1999) 503


\end{thebibliography}
\end{document}